\documentclass{aastex631}

\usepackage{soul}
\usepackage{booktabs}
\usepackage{array}

\begin{document}

\title{Identification of Likely Methane Absorption Features in the Optical Spectra of Titan}

\author[0000-0002-3392-6956]{Sirinrat Sithajan}
\affiliation{National Astronomical Research Institute of Thailand (Public Organization) \\
Don Kaeo, Mae Rim, Chiang Mai, 50180, Thailand}

\author[0009-0009-1992-8902]{Lalita Kaewbiang}
\affiliation{National Astronomical Research Institute of Thailand (Public Organization) \\
Don Kaeo, Mae Rim, Chiang Mai, 50180, Thailand}

\author[0000-0003-0433-3665]{Hugh R. A. Jones}
\affiliation{Centre for Astrophysics Research, University of Hertfordshire \\ College Lane, Hatfield AL10 9AB, UK}

\author[0000-0003-1916-9976]{Pakakaew Rittipruk}
\affiliation{National Astronomical Research Institute of Thailand (Public Organization) \\
Don Kaeo, Mae Rim, Chiang Mai, 50180, Thailand}

\author[0009-0006-2813-7799]{Sukanya Meethong}
\affiliation{National Astronomical Research Institute of Thailand (Public Organization) \\
Don Kaeo, Mae Rim, Chiang Mai, 50180, Thailand}

\begin{abstract}

The optical spectra of Titan reveal a rich set of absorption features, most of which are likely associated with methane (CH$_4$). Methane is a key molecule in planetary and exoplanetary atmospheres, yet a comprehensive high-resolution linelist at optical wavelengths remains incomplete. This study identified and characterized potential CH$_4$ absorption features in high-resolution optical spectra of Titan, providing essential data for linelist development and improving CH$_4$ detection and characterization. We analyzed Titan spectra from the ESPRESSO spectrograph (R~$\approx$~190{,}000), identifying intrinsic features and measuring their relative strengths. A conservative detection approach was employed, slightly overestimating solar and telluric contributions to distinguish them from Titan’s intrinsic features. To assess the impact of spectral resolution, we compared the ESPRESSO data with Titan UVES data (R~$\approx$~110{,}000). We identified 6{,}195 absorption features in the ESPRESSO spectra potentially associated with CH$_4$, of which 5{,}436 are newly reported. ESPRESSO detected twice as many features as UVES in overlapping regions, highlighting the advantage of higher-resolution data. Most detected lines remained unresolved, so our reported features are primarily blended absorption structures. We estimated the detection limit for feature identification to correspond to a CH$_4$ absorption coefficient of approximately 0.02~km-am$^{-1}$. Comparison of our results with a previous analysis of Titan UVES spectra and with experimental CH$_4$ data at a similar temperature showed good agreement, while some discrepancies were observed when compared with data acquired at a different temperature. We provide a comprehensive list of Titan absorption features with key reliability metrics, along with Titan’s intrinsic spectra, to support future studies.

\end{abstract}

\section{Introduction} 
\label{sec:intro}

In recent years, significant advancements in observational techniques and high-precision instruments have accelerated the detection and characterization of planetary bodies. As we explore a broader range of these bodies, the need for detailed knowledge of atmospheric composition, structure, and dynamics has become increasingly essential. Among various molecules, methane (CH$_4$) stands out due to its significance in planetary atmospheres. It is not only abundant in gas giants but also holds potential as a biomarker in smaller, rocky planets  (\citealt{Schwieterman_2018}; \citealt{Thompson_2022}).

Methane is the third most abundant molecule in the atmospheres of Jovian planets after hydrogen and helium, comprising about 0.2\% of the total molecular content \citep{2004jpsm.book...59T}. Its infrared radiation largely controls the stratospheric thermal profile. Thus, enhanced emissions are associated with upper atmospheric heating, like in the auroral regions \citep{KIM2015217}, or during the collision of the fragments of comet Shoemaker Levy 9 with Jupiter \citep{VladimirEFortov_1996}. Methane can also be used as a tracer of upper cloud variations, as a thermometer for retrieving the thermal profile, and as an indicator of the turbulence in the upper stratosphere, from the detection of fluorescence in methane bands \citep{2024A&A...692L..11P}.

Methane is particularly intriguing because of its dual origin: it is produced through abiotic processes such as serpentinization and volcanic outgassing, as well as biotic processes like methanogenesis on Earth \citep{2013ApJ...777...95S}. The detection of methane in the atmosphere of a rocky and temperate planet, especially when found alongside gases like oxygen or ozone, could suggest biological activity (\citealt{2013ApJ...775..104S}; \citealt{2018AsBio..18..630M}). Consequently, methane has become a primary focus in exoplanet atmospheric studies, particularly for rocky exoplanets that orbit within the habitable zones of their stars.

High-resolution spectroscopy has opened new possibilities in exoplanet atmospheric characterization (e.g., \citealt{2010Natur.465.1049S}; \citealt{Birkby_2013, Birkby_2017}; \citealt{Carleo_2022}). This powerful technique allows the identification of individual molecules in distant atmospheres by analyzing their unique spectral fingerprints. Through this approach, the chemical composition of exoplanet atmospheres can be determined, and details about their structure and dynamic processes, such as wind patterns and temperature variations, can be uncovered. 

Significant progress has been made in detecting and analyzing molecular features, including methane, in the infrared spectra of planetary atmospheres. However, spectra at shorter wavelengths remain largely underutilized for methane studies due to the scarcity of high-resolution absorption data (e.g., \citealt{Campargue_2023}). Molecular databases such as HITRAN \citep{GORDON20173}, HITEMP \citep{ROTHMAN20102139}, ExoMol \citep{TENNYSON201673}, and TheoReTS \citep{REY2016138} provide extensive methane linelists in the infrared, but their coverage is significantly limited at shorter wavelengths (\citealt{REY2018114}; \citealt{Hargreaves_2020}; \citealt{10.1093/mnras/stae148}). Developing a comprehensive methane linelist at these wavelengths is particularly challenging due to the intrinsically weak absorption associated with high overtone and combination bands. In addition, these transitions exhibit notable temperature dependence, as demonstrated in near-infrared studies such as \cite{CAMPARGUE2012110} and \cite{DEBERGH201285}, further complicating the construction of an accurate linelist. 

Accurate modeling of methane absorption in planetary atmospheres requires not only accurate line positions and intensities, but also reliable lower-state energy values, which determine how intensities change with temperature. Such models are essential for the reliable interpretation of planetary atmospheres. The current scarcity of a high-resolution linelist at optical wavelengths is especially relevant for upcoming high resolution spectrographs such as RISTRETTO \citep{Lovis_2024} on the Very Large Telescope (VLT) and ANDES \citep{marconi2024} on the Extremely Large Telescope (ELT), which aim to detect starlight reflected from exoplanetary atmospheres, where much of the light lies in the optical wavelength range.

To enhance the incomplete methane linelist in the optical wavelengths, Titan serves as a promising natural laboratory. Its cold ($T \approx 100\,\mathrm{K}$) and dense atmosphere provides long atmospheric pathlengths that facilitate the detection of the weak high-overtone transitions at low temperatures. These transitions are otherwise difficult to observe under typical Earth-based laboratory conditions. Titan's atmosphere is composed primarily of molecular nitrogen (N\textsubscript{2}; over 90\%) and CH\textsubscript{4} (less than 10\%, varying with altitude), along with trace constituents that together account for less than 1\% of the total atmospheric content. These minor constituents include hydrocarbons such as ethane (C$_2$H$_6$), propane (C$_3$H$_8$), ethylene (C$_2$H$_4$), and tricarbon (C$_3$), as well as molecular hydrogen (H$_2$) \citep{Nixon_2024}. In the optical wavelength range, Titan’s molecular absorption is dominated by CH$_4$ features, while contributions from minor species are expected to be extremely weak (e.g., \citealt{KARKOSCHKA1994174, KARKOSCHKA1998134, KARKOSCHKA2010674, RIANCOSILVA2024105836}). This is attributed to both their low abundances and the intrinsically weak nature of their optical transitions. For instance, the hydrocarbons are typically present at concentrations ranging from parts per million to parts per billion, and their optical features arise primarily from high overtone and combination bands. Such bands are significantly weaker than their fundamental vibrational transitions, which typically occur in the infrared, or electronic transitions, which generally occur in the ultraviolet. 

One notable exception is the C$_3$ molecule, which exhibits relatively strong electronic transitions near 4050~\AA. These transitions are intrinsically stronger than vibrational overtones and therefore have the potential to produce distinct absorption signatures in Titan’s optical spectrum. However, its very low abundance, combined with the increasing opacity of Titan’s haze at shorter wavelengths \citep{MCKAY2001}, poses a significant challenge for detection (e.g., \citealt{RIANCOSILVA2024105836}). Although contributions from trace species cannot be entirely ruled out, especially in ultra high-resolution spectra (e.g., \citealt{2024Rukdee}), the vast majority of absorption features detected in Titan’s visible spectra in this study are expected to originate from methane.

Advances in spectroscopic instruments, such as the Echelle SPectrograph for Rocky Exoplanet and Stable Spectroscopic Observations (ESPRESSO; \citealt{2021A&A...645A..96P}),
now provide unprecedented signal-to-noise ratios (SNRs) and precision for studying methane’s visible spectrum. By utilizing data from ESPRESSO, along with a lower-resolution optical spectrograph such as the Ultraviolet and Visual Echelle Spectrograph (UVES; \citealt{2000SPIE.4008..534D}), we aim to identify and characterize high-resolution methane absorption features in this range.

In this paper, Section~\ref{sec:method} describes the data and method. Section~\ref{sec:results} presents the identification and strength measurements of methane features in Titan's spectrum. Section~\ref{sec:resolutions} examines the impact of spectral resolution on the observed methane features. Section~\ref{sec:compare} compares our findings with those from previous studies, and Section~\ref{sec:summary} summarizes the key conclusions.

\section{Data and Method} 
\label{sec:method}

\subsection{Data} \label{subsec:data}
We analyze three spectra of Titan obtained from the European Southern Observatory (ESO) Science Archive\footnote{\url{https://archive.eso.org/wdb/wdb/adp/phase3_spectral/form}}, processed using standard ESO pipelines. The spectra were recorded on July 21, 2021, in the Ultra High Resolution (UHR) mode of ESPRESSO (R $\approx$ 190,000; $\Delta\lambda$ = 3772–7900 \AA; \citealt{2021A&A...645A..96P}), installed on the VLT at Paranal Observatory, Chile. Each spectrum was acquired with an exposure time of 16 minutes. These observations were conducted under program ID 106.218L.001 (PI: Turbet, M.), and a summary is provided in Table~\ref{tab:observations}.

Titan’s angular size at the time of observation was estimated to be approximately 0.7\arcsec, based on ephemeris data from NASA JPL\footnote{\url{https://ssd.jpl.nasa.gov/planets/eph_export.html}} and \cite{Park_2021}, which is comparable to the 0.5\arcsec diameter of ESPRESSO's fiber. This suggests that the spectra likely captured the entire disk of Titan.

\begin{table*}
\centering \caption{Details of the spectral data of Titan used for methane feature identification and characterization. The median SNR is calculated over the wavelength range $4000\text{--}7870 \, \text{\AA}$ for ESPRESSO data and $4170\text{--}6210 \, \text{\AA}$ for UVES data. The last column shows the Doppler correction applied to shift the spectra to Titan’s rest frame from the Solar System barycenter. The correction values correspond to the mid-exposure time of each observation.}
\label{tab:observations}
\begin{tabular}{l|c|c|c} 
\hline
 \textbf{Instrument/Telescope} & \textbf{Date and Time of Observation (UT)} & \textbf{Median SNR} &
 \textbf{Doppler Correction (km/s)} \\
\hline\hline
ESPRESSO/VLT   & 2021-07-21 04:07:00.852 &  170 & -5.177 \\
 & 2021-07-21 04:23:44.485 & 165 & -5.188 \\
 & 2021-07-21 04:40:29.289 & 163 & -5.199 \\
\hline
UVES/VLT & 2002-09-08 08:56:44.824        &   138  & 4.331 \\
     & 2002-09-08 09:16:55.120       &   139  & 4.346 \\
     & 2002-09-11 09:23:19.835       &   144  & 3.791 \\
     & 2002-09-22 08:42:05.549       &   143  & 1.136 \\
      & 2002-09-22 09:02:09.260       &   146  & 1.164 \\
      & 2002-11-07 07:07:04.239       &   158  & -2.629 \\  
      & 2002-11-07 07:33:11.557       &   158  & -2.600 \\    
      & 2002-12-12 04:34:11.344    &  158  & 2.940 \\
       & 2002-12-12 04:54:18.693       &   156  & 2.963 \\
      & 2002-12-12 05:23:05.229       &   158  & 2.995 \\
      & 2002-12-12 05:43:10.306       &   159  & 3.020 \\
      & 2002-12-12 06:05:14.700       &   157  & 3.045 \\
      & 2002-12-12 06:25:19.389       &   150  & 3.068 \\
      & 2003-01-02 04:29:56.514       &   164  & 2.296 \\
      & 2003-01-02 04:50:15.247       &   161  & 2.270 \\
\hline
\end{tabular}
\end{table*}

\subsection{Identification of Methane Features} \label{subsec:featureid}
The three Titan spectra are normalized using the RASSINE (Rolling Alpha Shape for a Spectrally Improved Normalisation Estimation; \citealt{Cretignier_2020}) algorithm. RASSINE is an open-source Python code\footnote{\url{https://github.com/MichaelCretignier/Rassine_public}} designed for normalizing merged 1D spectra, such as those used in this work. It models the upper envelope of a spectrum using convex-hull and alpha-shape theories, based on the assumption that this envelope traces the continuum. This approach is effective for solar-type stars but may break down in regions with dense spectral lines, such as the blue end of the visible spectrum or in M dwarfs, where molecular bands dominate.

\begin{figure*}[ht]
    \centering   \includegraphics[width=\textwidth]{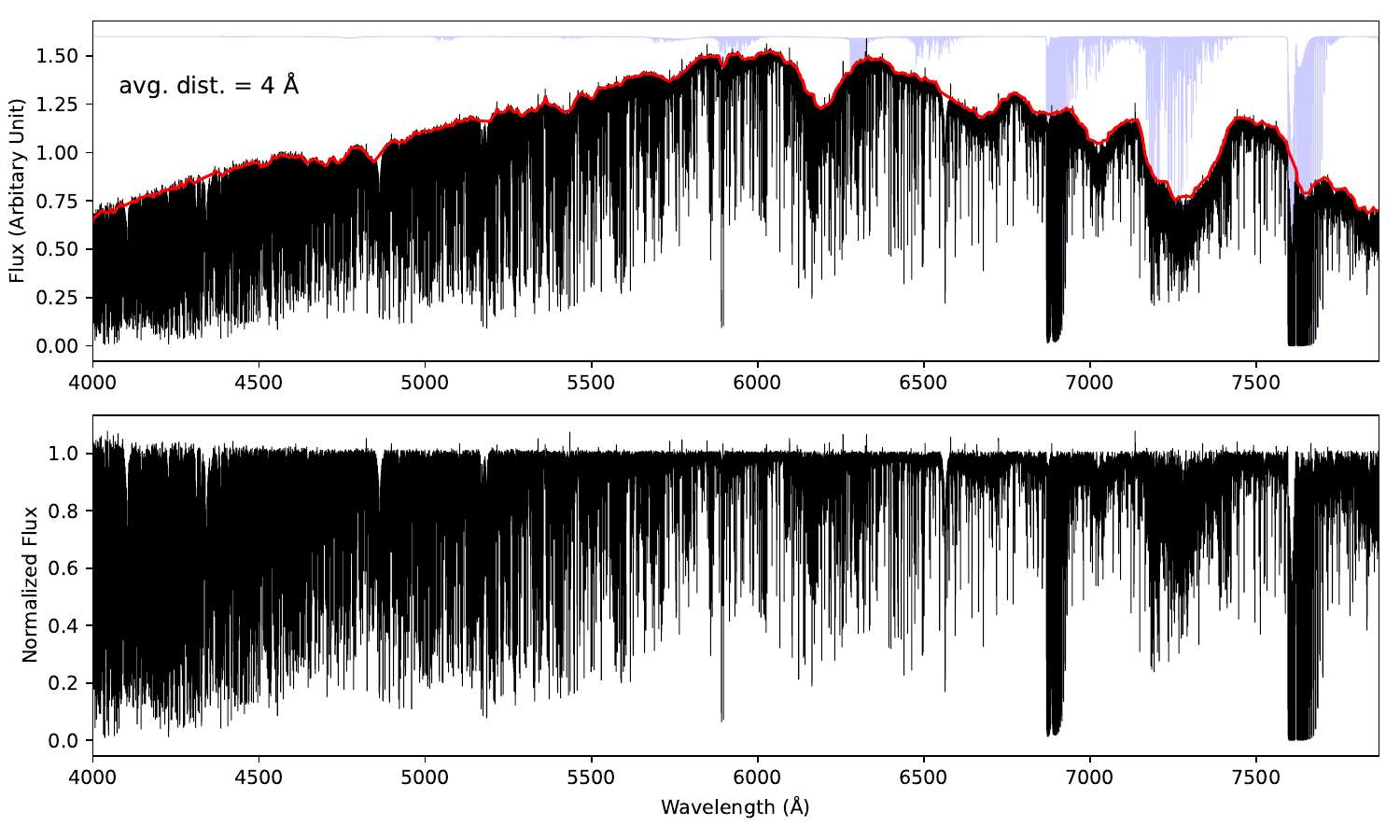} \caption{ESPRESSO spectrum of Titan before (top) and after (bottom) continuum normalization with RASSINE. The red curve in the top panel represents the estimated continuum level, constructed by linearly connecting the anchor points (not shown in this plot) described in the text. The average distance between anchor points is approximately 4 \AA, as indicated by the label on the left edge of the top panel. The blue line in the top panel serves as a reference for telluric absorption across different wavelengths.}
    \label{fig:normalization}
\end{figure*}

To avoid subjective continuum placement and ensure reproducibility, we run RASSINE in automatic mode, specifying fixed parameters where necessary. These include predefined regions containing broad solar features, such as the Balmer series and the MgIb triplet, as well as the parameter \texttt{par\_reg\_nu = poly\_1.0}, which is not determined automatically by the algorithm. In automatic mode, RASSINE determines most key input parameters from the average absorption line width computed directly from the spectrum. For the ESPRESSO Titan spectra, which are dominated by solar absorption features, this width is approximately 0.1~\AA. This value, together with the overall structure of the spectrum, governs the selection of `anchor points', defined as a subset of local maxima that are connected to construct the continuum. Figure~\ref{fig:normalization} shows one of the three Titan spectra before and after continuum normalization with RASSINE. The continuum, shown as a red curve, is constructed by linearly connecting the anchor points identified by the algorithm.

\begin{figure*}[ht]
    \centering   \includegraphics[width=\textwidth]{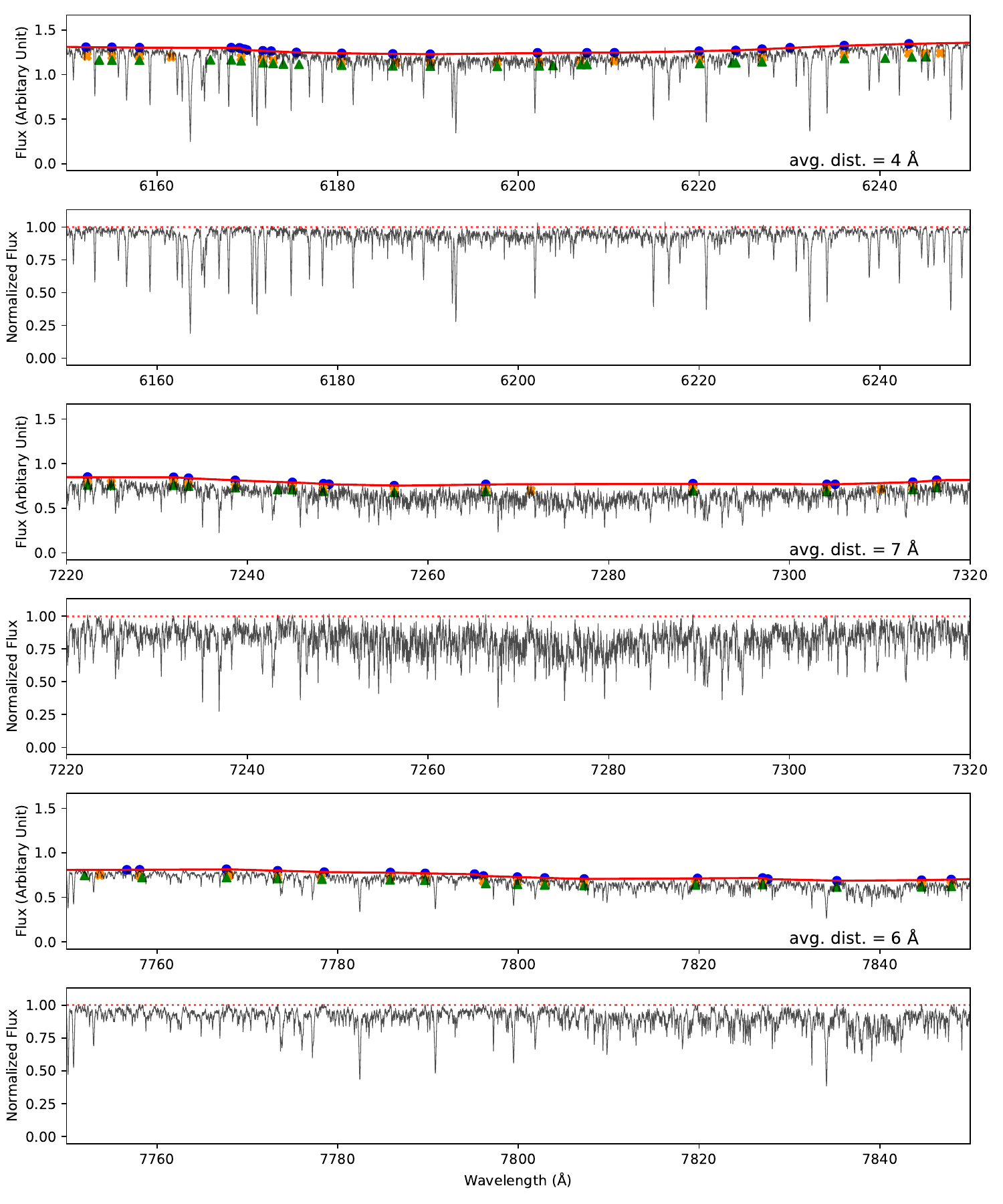} \caption{Same spectrum as shown in Figure~\ref{fig:normalization}, but zoomed into 100\,\AA-wide segments centered on selected regions: (rows 1--2) regions with moderate CH$_4$ absorption, (rows 3--4) regions with strong absorption from both CH$_4$ and telluric features, and (rows 5--6) regions dominated by strong CH$_4$ absorption alone. The red solid line indicates the continuum level from Figure~\ref{fig:normalization}, with blue dots marking its anchor points. Orange crosses and green triangles denote the anchor points from the remaining two spectra. A red dashed line at a normalized flux of 1 marks the baseline of the normalized spectrum. The average distance between neighboring anchor points in each region is shown in the lower right corner of the corresponding panel.}
    \label{fig:normalization_regions}
\end{figure*}

Figure~\ref{fig:normalization_regions} illustrates how anchor points vary across the three spectra of Titan and among different spectral regions within each spectrum. The black and red lines are the same as those shown in Figure~\ref{fig:normalization}, with anchor points for that specific spectrum marked by blue circles. Anchor points from the remaining two spectra are indicated by orange crosses and green triangles. Each panel displays selected regions, including those with moderate CH$_4$ absorption, regions with strong absorption from both CH$_4$ and telluric features, and regions dominated by strong CH$_4$ absorption alone. The average spacing between anchor points is indicated for each region. This figure demonstrates that the distribution of anchor points varies across the spectra, despite the use of identical manual input settings. These differences arise solely from the intrinsic structure of each observed spectrum. The variation in anchor point placements and the resulting continuum reflects the precision of spectral normalization. Furthermore, the average spacing between anchor points can be taken as a rough measure of the maximum feature width below which spectral features retain their strength after normalization. Features broader than this threshold are likely to have their strength reduced or be suppressed during the continuum normalization process. The average spacing is approximately 4~\AA\ for the three spectra analyzed for this work.

In this study, we do not perform manual continuum adjustments, except for the removal of a few obvious outliers, such as those caused by cosmic rays. This choice is motivated by several complicating factors that affect the underlying continuum, including residual instrumental response, time-dependent atmospheric extinction, and the blending of dense CH$_4$ lines, as discussed in later sections.

Since the spectra were acquired consecutively within a short time frame and the ESPRESSO instrument exhibits excellent stability, the primary sources of flux variation in the normalized spectra are photon noise and the precision of the normalization process. We trim the data to a wavelength range of 4000–7870 \AA\ to exclude regions with low SNRs, particularly near the spectral edges, especially in the blue wavelengths, where the SNR falls below 20. After trimming, the median flux variation across the three normalized spectra is approximately 0.6\%.

The spectra are shifted to Titan's rest frame using NASA JPL ephemerides, with the shift values provided in the last column of Table~\ref{tab:observations}. This process aligns Titan's intrinsic spectral features while leaving the solar and Earth's atmospheric (telluric) components slightly misaligned. However, because the Titan spectra were acquired within a short time window, the misalignment of solar and telluric features is minimal. The three spectra are combined to enhance the SNR for methane feature identification, yielding a final average SNR of 278 for the combined Titan spectrum.

Each Titan spectrum generally consists of three main components: solar, telluric, and intrinsic Titan features, the latter primarily expected to be CH$_4$ absorption. To identify these potential CH$_4$ features, we construct a ``background" spectrum consisting of the corresponding solar and telluric components of the combined Titan spectrum. This background serves as a reference, with CH$_4$ features identified as those that stand out significantly above the background. For the solar component, we use the high-resolution ``Solar Flux Atlas from 296 to 1300 nm" \citep{1984sfat.book.....K} from the NSO historical archive\footnote{\url{https://nso.edu/data/historical-archive}}, obtained with the Fourier Transform Spectrometer (FTS) at the McMath-Pierce Solar Facility on Kitt Peak, Arizona. This spectrum, with a resolving power of R $\approx$ 480,000, is convolved with a Gaussian profile to match the resolution of the ESPRESSO spectrum and then Doppler-shifted to align with the solar lines in the Titan spectrum.

Since the solar spectrum contains telluric contamination that could obscure CH$_4$ features, we apply a partial correction to enhance their visibility. We carefully avoid overcorrection, as it can introduce artifacts that might lead to misidentifying telluric residuals as CH$_4$ features. To minimize this risk, we reduce the strength of telluric lines rather than fully correcting them. Our focus is on partially correcting H$_2$O and O$_2$ lines, while other telluric molecules produce lines too weak to significantly impact the analysis and are therefore left uncorrected. Hereafter, unless otherwise specified, “telluric lines” will specifically refer to H$_2$O and O$_2$ absorption lines. This partial correction employs a telluric model generated with TAPAS\footnote{\url{https://tapas.aeris-data.fr/}} \citep{2014A&A...564A..46B}, where the telluric lines are modeled to be 10\% weaker than those observed in the solar spectrum, with about 95\% of lines meeting this reduction criterion. It is important to note that telluric lines in the solar spectrum and their counterparts in the model do not always scale uniformly. This approach ensures that most lines in the model remain weaker than those in the solar spectrum, reducing the likelihood of misidentifying correction residuals as CH$_4$ features.

\begin{figure*}[ht]
    \centering   \includegraphics[width=\textwidth]{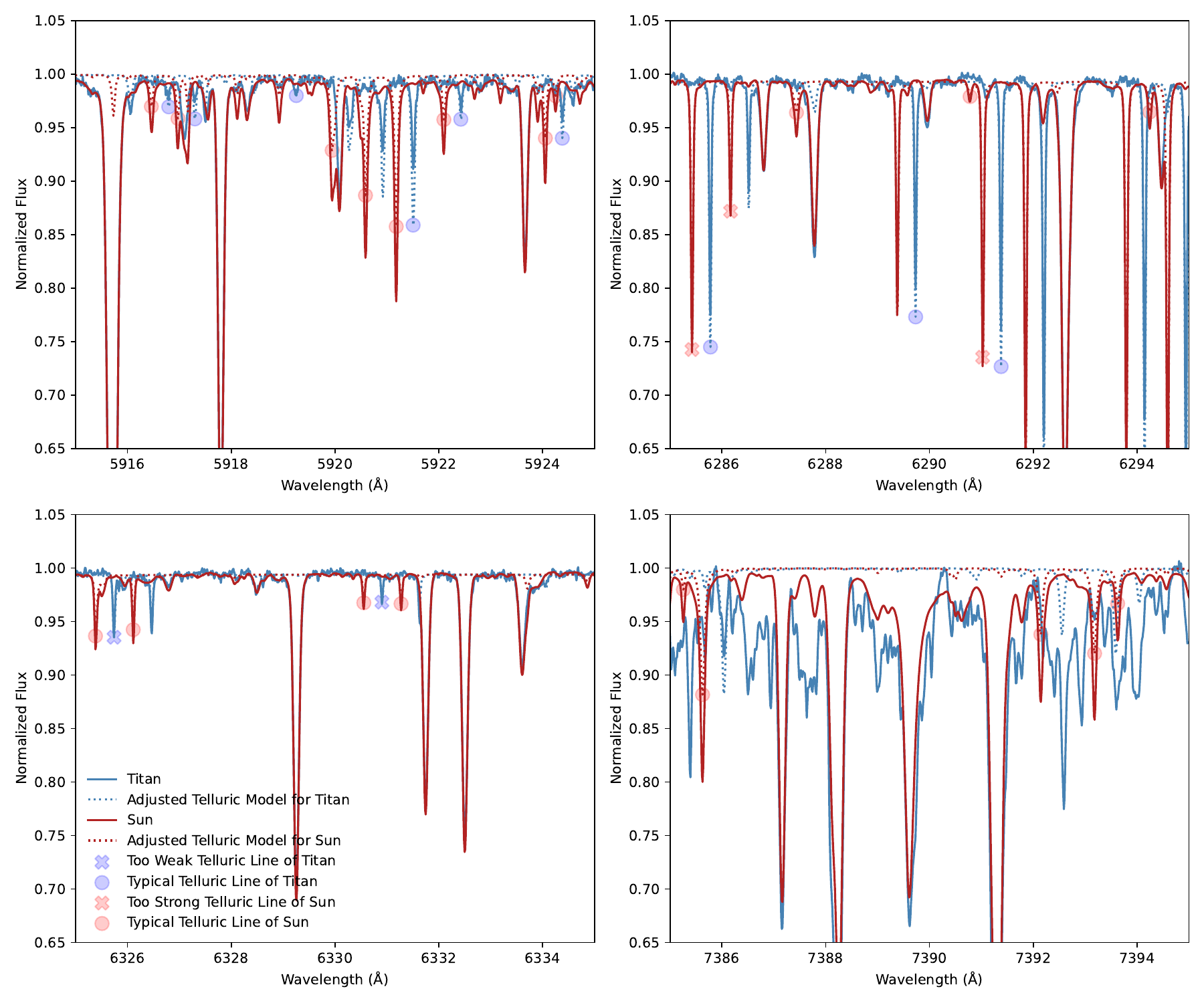} \caption{An illustration of the adjusted telluric models for the solar and Titan spectra, where a 10\% reduction from the best-fit model is applied to the solar spectrum and a 10\% increase from the best-fit model is applied to the Titan spectrum, as described in Section \ref{subsec:featureid}. `o' and `x' markers indicate uncontaminated telluric lines used to estimate these adjustments. `o' markers represent lines with expected strength, while `x' markers denote lines with anomalous strength. The center of each `o' or `x' corresponds to the peak of the adjusted telluric model. The four regions shown were selected to clearly demonstrate both marker scenarios and to span a broad wavelength range relevant to the study.}
    \label{fig:puretelluric}
\end{figure*}

For the telluric component of the background, a model is generated in which the telluric lines are 10\% stronger than those in the Titan spectrum, with approximately 95\% of the lines meeting this criterion. Figure \ref{fig:puretelluric} illustrates the adjusted telluric models for the solar and Titan spectra, aiding in the differentiation of Titan’s intrinsic features from telluric contamination. In the figure, `o' markers denote uncontaminated telluric lines with typical strength, while `x' markers indicate those with anomalous strength, emphasizing that model lines do not always scale uniformly to observations. These uncontaminated lines are used to adjust telluric model strength. For the solar spectrum, we identify uncontaminated telluric lines by referencing solar line locations from a synthetic spectrum generated with the SYNTHE stellar model synthesis code \citep{2005MSAIS...8...14K}. This synthetic spectrum is computed based on the resolving power of FTS McMath data, the Sun's effective temperature of 5777 K, $\log g = 4.437$, the abundances from \citet{2021A&A...653A.141A}, and atomic and molecular data from the Kurucz database. For the Titan spectrum, uncontaminated telluric lines are identified by referring to the observed solar spectrum and CH$_4$ candidate feature locations.

Modeling both weaker and stronger telluric lines, as described above, helps establish that the detected features are genuine CH$_4$ features originating from Titan’s atmosphere. A trade-off of this method is that some CH$_4$ features overlapping with telluric lines may be missed, especially those weaker than the telluric residuals from the solar spectrum or only slightly stronger than the telluric lines in the Titan spectrum. However, when measuring the relative strength of CH$_4$ features in Section~\ref{subsec:featuredepth}, we use telluric models that best match the observed telluric lines in both the solar and Titan spectra to obtain more accurate strength measurements.

\begin{figure*}[ht]
    \centering    \includegraphics[width=\textwidth]{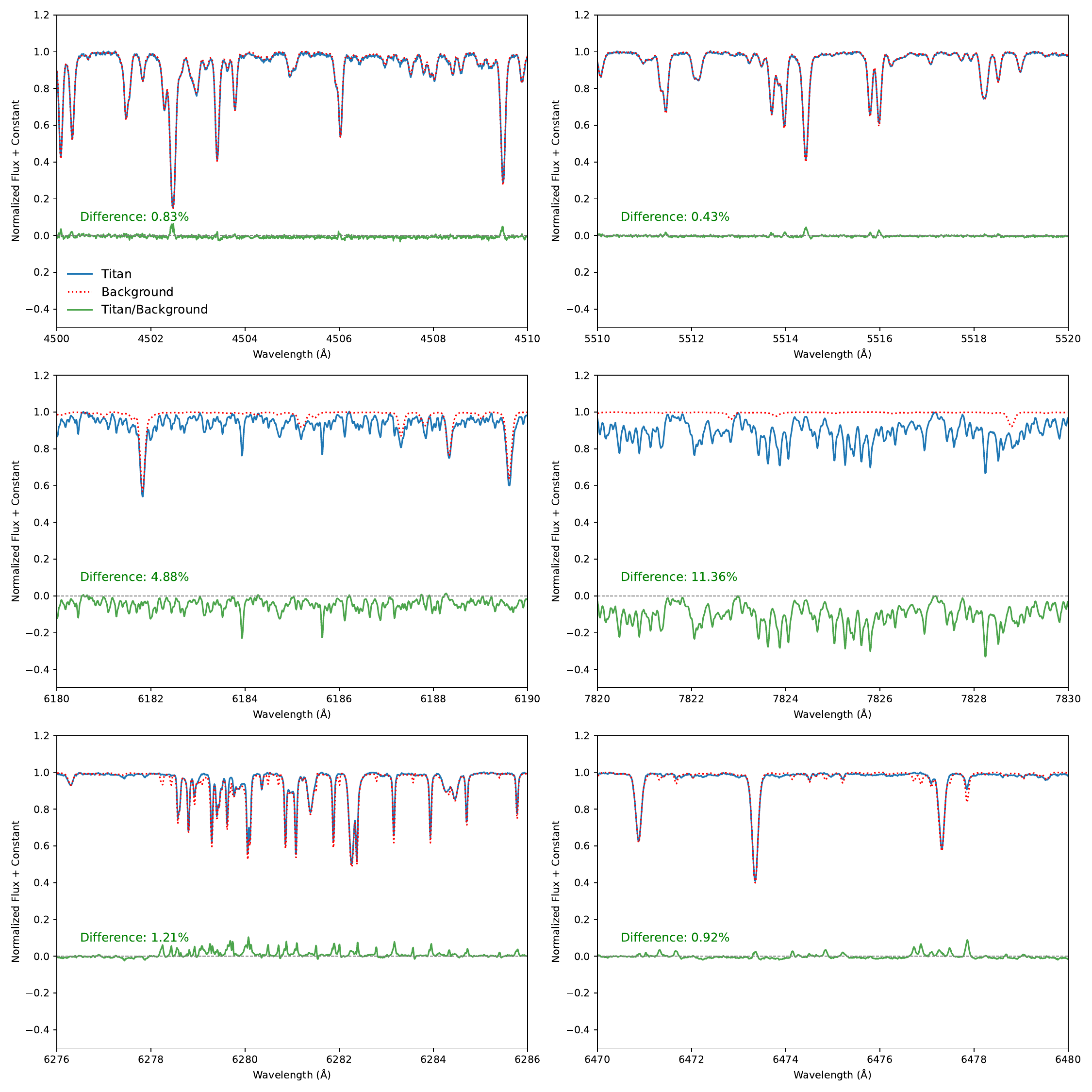} \caption{Comparison of the Titan spectrum with the  ``background":  (Top) Regions dominated by solar lines. (Middle) Regions with minimal telluric and solar interference. (Bottom) Regions with prominent telluric O$_2$ and H$_2$O lines. Telluric and solar features in the Titan spectrum align with their counterparts in the background, while telluric residuals from the solar spectrum appear only in the background, without overlap in the Titan spectrum. Features not originating from Titan typically produce positive residuals, except near the wings of strong lines, where negative residuals may occur. The percentage difference between the Titan spectrum and the background, providing a measure of their variation, is calculated by summing the absolute differences and dividing by the average of the two spectra.}
    \label{fig:titan_bg}
\end{figure*}

The background spectrum is normalized using the RASSINE algorithm to ensure consistency with the Titan spectrum. The Titan spectrum is then divided by the background, producing a residual spectrum used to identify CH$_4$ features. Figure \ref{fig:titan_bg} displays the Titan spectrum, the background, and the residual spectrum across selected regions. The top panels highlight regions dominated by solar lines, where the Titan and background spectra are nearly identical except near strong solar lines. In these areas, the background appears slightly deeper, producing positive residual peaks. The middle panels focus on regions with minimal telluric and solar interference, where CH$_4$ features dominate and appear as negative residual peaks. The bottom-left and bottom-right panels show regions with prominent O$_2$ and H$_2$O lines, respectively, which also generate positive residual peaks. Across all regions, small negative residuals occasionally appear near the wings of strong solar and telluric lines or along the edges of positive residuals, potentially leading to misidentifications of CH$_4$ features in our automated detection process. To mitigate this issue, we include a cautionary note, labeled `ST mask', for CH$_4$ features identified in regions near these strong lines (absorption $>$ 30\%).

An absorption peak in the residual spectrum is classified as a CH$_4$ feature if its strength exceeds the local noise by at least a factor of five. The local noise is estimated from variations averaged over a 2 \AA\ width among the three individual residual spectra, obtained by dividing the three Titan spectra by the background. Note that we avoid referring to it as a CH$_4$ absorption ``line" because, at the resolving power of ESPRESSO ($R \approx 190{,}000$), the observed CH$_4$ absorption features are mostly unresolved, and individual lines cannot be distinctly separated.

\subsection{Measurement of the Relative Strength of Methane Features}  \label{subsec:featuredepth}

After obtaining a list of CH$_4$ features, we measure their relative strengths, defined as the depth of each absorption peak relative to the ``best-fit background". This best-fit background is generated using the approach outlined in Section~\ref{subsec:featureid}, but it employs telluric models that most closely match the observed telluric lines, rather than models with weaker or stronger telluric features. The estimated difference between the best-fit telluric models and the observed telluric lines is approximately 1\% for the solar spectrum and about 2\% for the Titan spectrum.

\begin{figure*}[ht]
    \centering    \includegraphics[width=\textwidth]{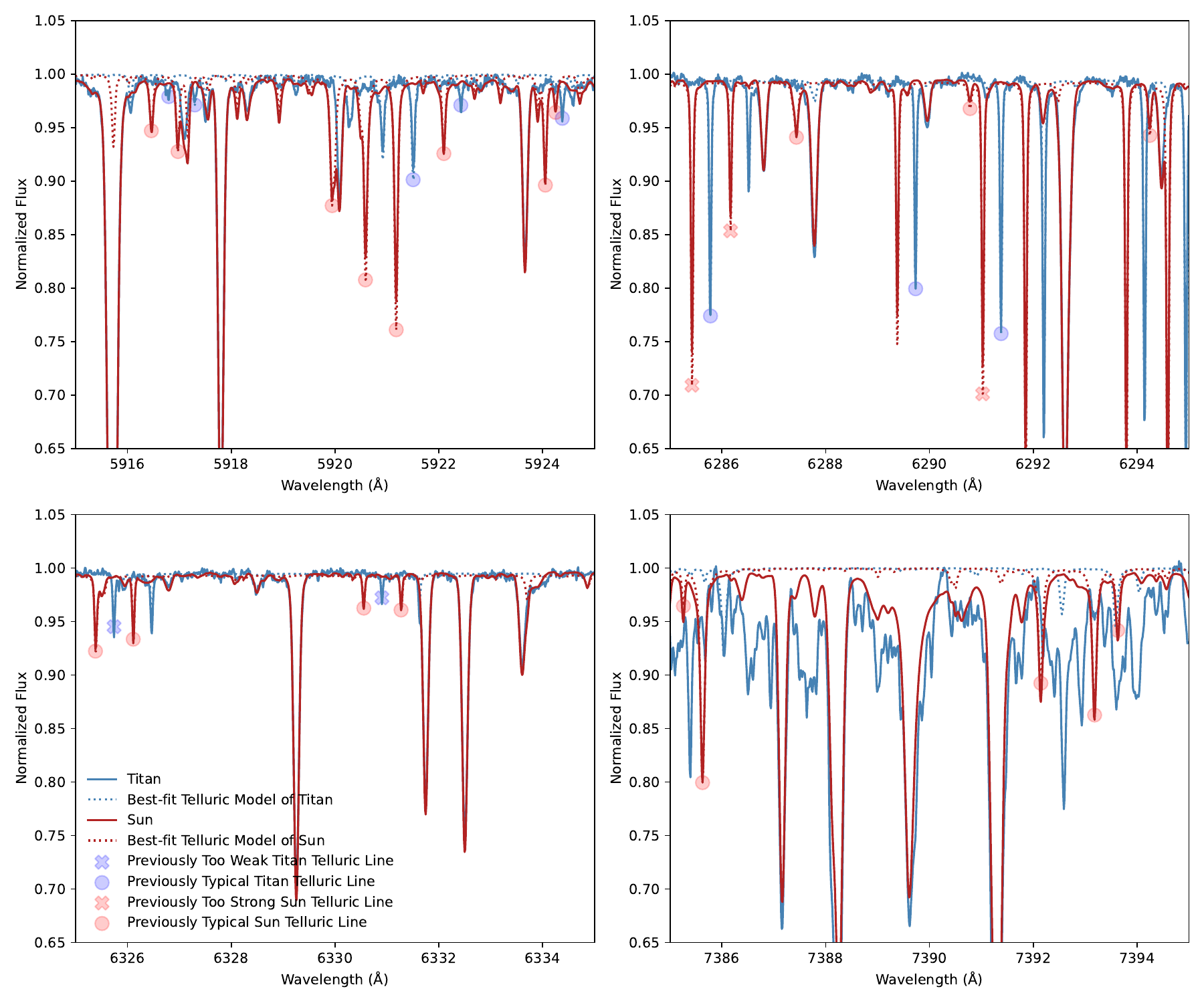} \caption{The best-matched telluric models for the observed telluric lines in both the solar and Titan spectra, in contrast to the adjusted models shown in Figure \ref{fig:puretelluric}. Unlike Figure \ref{fig:puretelluric}, where a 10\% reduction and increase were applied to the best-fit models for the solar and Titan spectra, respectively, the models here represent the closest match to the observed telluric features without adjustments. The `x' and `o' markers indicate the same wavelength positions as in Figure \ref{fig:puretelluric}, with their vertical positions adjusted to align with the peak centers of the best-fit telluric model lines. This alignment allows for a direct comparison of line depths relative to the weaker and stronger models illustrated in Figure \ref{fig:puretelluric}.}
    \label{fig:puretelluric_bfbg}
\end{figure*}

\begin{figure*}[ht]
    \centering    \includegraphics[width=\textwidth]{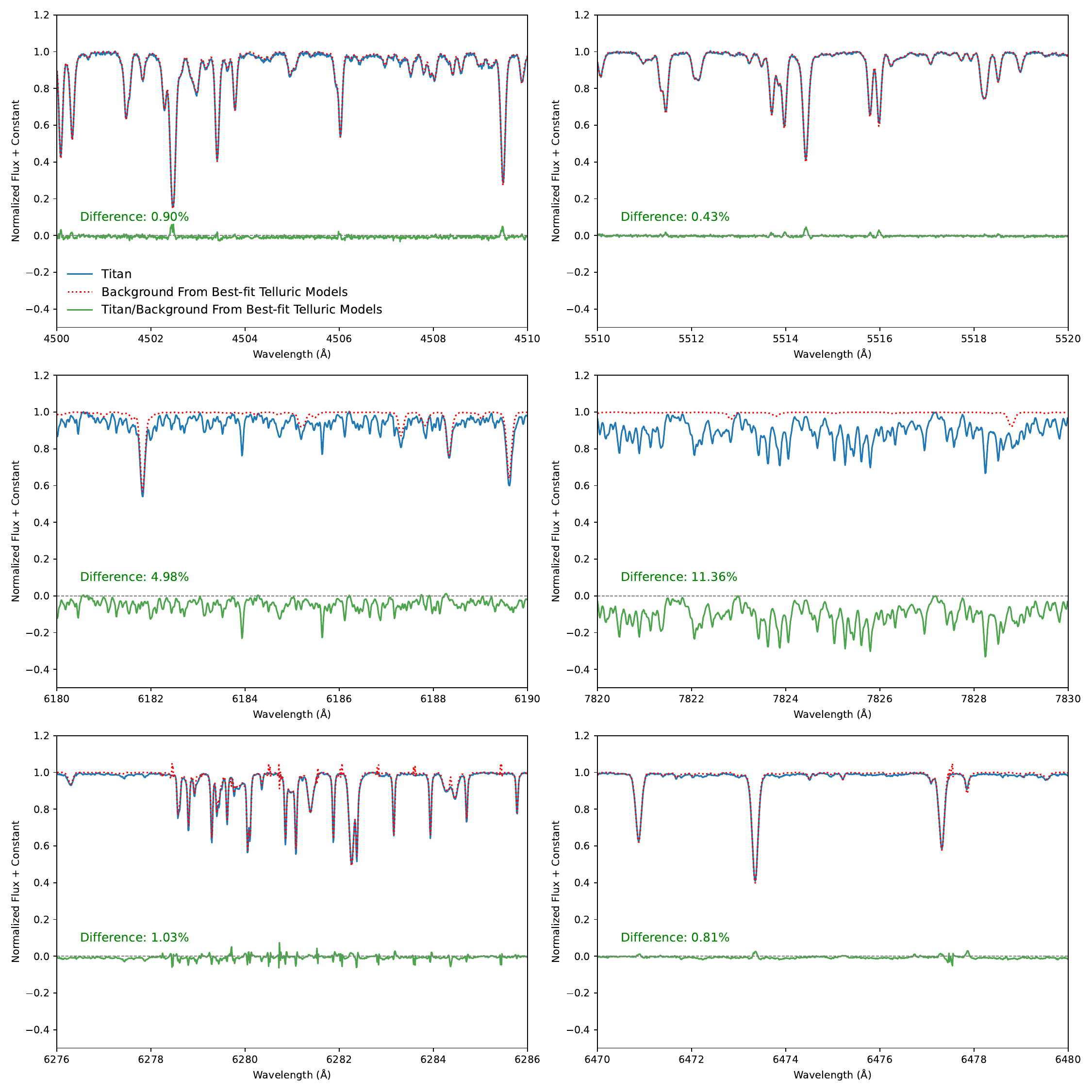} \caption{Comparison of the Titan spectrum with the ``best-fit background." The panel structure is similar to that in Figure \ref{fig:titan_bg}, but here the background is constructed using telluric models that best match the observed telluric lines in both the solar and Titan spectra, referred to as the ``best-fit background." In this case, negative residuals are no longer solely attributed to Titan's absorption.}
    \label{fig:titan_bfbg}
\end{figure*}

Figure \ref{fig:puretelluric_bfbg} presents plots similar to those in Figure \ref{fig:puretelluric}, but with the previously adjusted telluric models replaced by the best-fit telluric models. The wavelength positions of the `x' and `o' symbols remain consistent with those in Figure \ref{fig:puretelluric}, but the peak centers have been adjusted along the vertical axis to align with the best-fit telluric models. This adjustment highlights differences in line depths compared to Figure \ref{fig:puretelluric}. For instance, in Figure \ref{fig:puretelluric}, the first red `x' in the top-right panel, located blueward of 6286~\AA, represents an adjusted telluric line model for the solar spectrum that is anomalously strong (although still weaker than the observed line). Under the best-fit telluric model in Figure \ref{fig:puretelluric_bfbg}, this line becomes much stronger than the observed line. Figure \ref{fig:titan_bfbg} is similar to Figure \ref{fig:titan_bg}, but it uses the best-fit background instead of the background from the feature identification step. This figure demonstrates that the best-fit telluric models can effectively remove non-Titan-origin features in some regions (e.g., the bottom-right plot around 6474–6476~\AA) but may also introduce artifacts that resemble CH$_4$ absorption features, potentially leading to false CH$_4$ identifications (e.g., the bottom-left plot). Together, these four figures illustrate the rationale for employing two distinct backgrounds for feature identification and relative strength measurement.

We identify CH$_4$ features using the same criteria outlined in Section~\ref{subsec:featureid} to compare results obtained with different backgrounds. This approach helps mitigate artifacts in the CH$_4$ feature list from Section~\ref{subsec:featureid}, which initially contained 6,273 (6,195+78) features. Using the best-fit background, we identify 6,672 (6,195+477) CH$_4$ features. By comparing the two lists, CH$_4$ features are classified into three groups: 1) 6,195 features common to both the initial CH$_4$ feature list and the best-fit background list, 2) 477 features unique to the best-fit background list, and 3) 78 features from the initial list that are absent in the best-fit background list.

\begin{figure*}[ht]
    \centering
    \includegraphics[width=\textwidth]{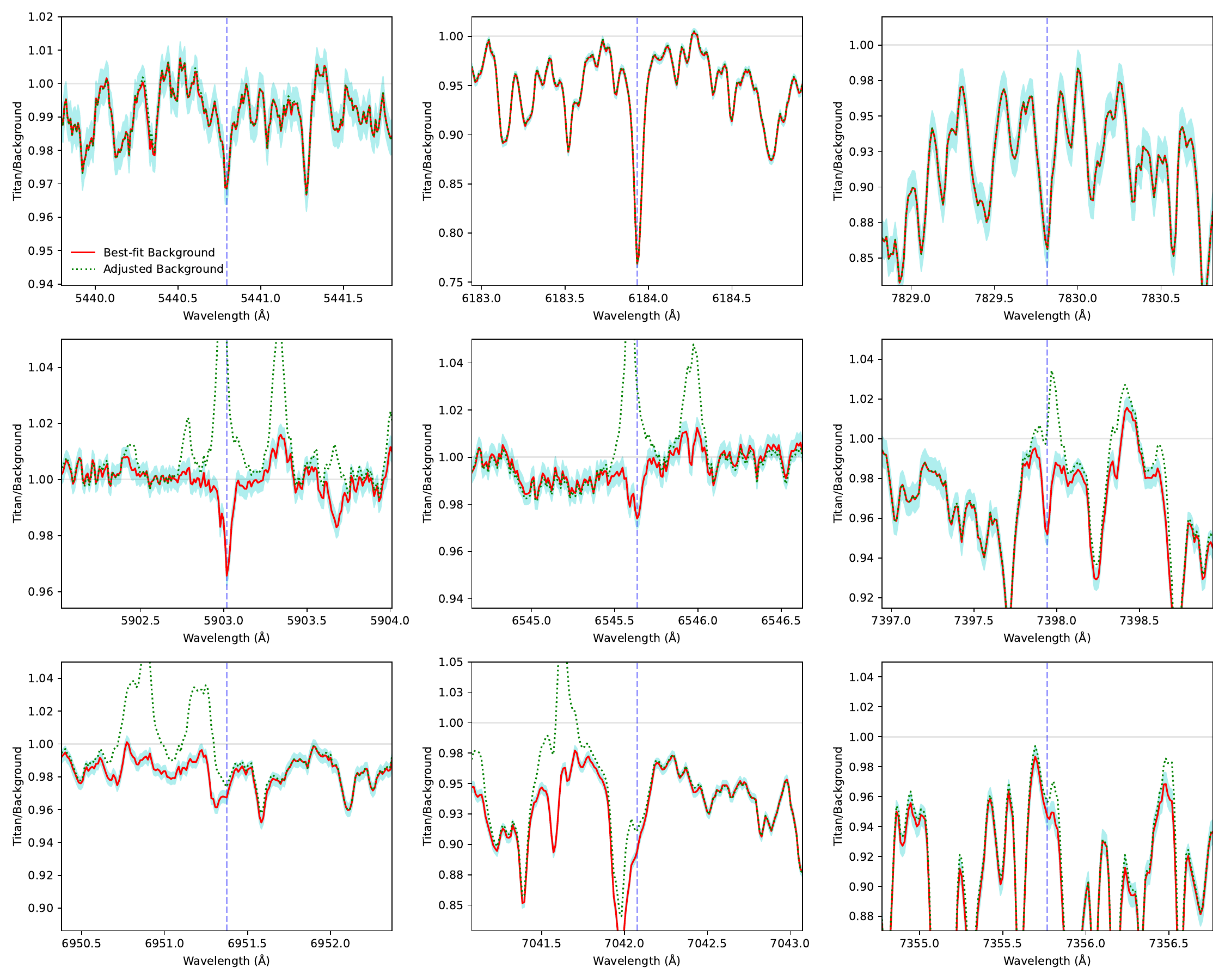} \caption{Examples of CH$_4$ features across the three categories discussed in Section~\ref{subsec:featuredepth}. (Top) Features detected in both the adjusted and best-fit backgrounds, indicating consistent detections. (Middle) Features unique to the best-fit background, i.e., peaks found in the red spectrum but not in the green-dot spectrum, suggesting that the improved background reveals these features, which could represent either genuine CH$_4$ features or artifacts. (Bottom) Features unique to the adjusted background, i.e., peaks found in the green-dot spectrum but not in the red spectrum, indicating that these features are sensitive to the choice of background and may not be robust. In each panel, the feature under discussion is marked with a vertical blue dashed line. Cyan bands represent variations across the three individual Titan spectra, and horizontal lines at $y = 1$ denote the background levels.}
    \label{fig:3categories}
\end{figure*}

The 477 features unique to the best-fit background list may correspond to weak features that become detectable with a more accurate background or to artifacts arising from overcorrected telluric lines in the solar spectrum or undercorrected telluric lines in the Titan spectrum. While some of these features are likely genuine Titan absorption features based on visual inspection, we refrain from adding them to the initial CH$_4$ list to minimize the risk of misidentifications based solely on visual assessment. The absence of 78 features from the initial list in the best-fit background case was initially unexpected. Further investigation revealed that these features were not detected in the best-fit scenario due to interference from the best-fit background, which reduced their peak prominence and altered their shapes. Given the relatively small number of features in this category and to maintain a conservative CH$_4$ list, we decided to exclude them from the initial CH$_4$ list, as they are particularly sensitive to background variations. This process resulted in a final list of 6,195 CH$_4$ absorption features. Example plots from all three categories are shown in Figure~\ref{fig:3categories}. We note that despite considerable modifications to the background, there was no dramatic change in the number of CH$_4$ features identified ($<$10\%). Therefore, we consider our list of CH$_4$ features to be relatively conservative.

The relative strength of a methane feature is determined from the depth of its absorption peak in the combined Titan spectrum, divided by the best-fit background. This combined Titan spectrum, with the best-fit background removed, is hereafter referred to as the Titan “intrinsic” spectrum. It is provided as supplementary material in Appendix\textbf{~\ref{sec:appendixa}.} The uncertainty in the relative strength is estimated from the difference between the maximum and minimum strengths measured across the three individual Titan spectra, each referenced to the best-fit background. 

We refer to this measurement as the “relative strength” rather than simply “strength” because it quantifies the prominence of absorption features above the underlying broad structures. These broad structures may arise from non-Titan sources, such as the telluric O$_2$ band near 5750~\AA, or from Titan itself, such as unresolved CH$_4$ absorption lines, as described by \citet{Campargue_2023}. Such structures typically span tens of angstroms and are therefore suppressed by the normalization process, which employs anchor points spaced, on average, every 4~\AA, as explained in Section~\ref{subsec:featureid}. In contrast, the individual CH$_4$ lines in the studied region are much narrower and densely packed, often spaced more closely than the resolution limit of ESPRESSO. As a result, over narrow wavelength intervals, such as a few angstroms, these CH$_4$ features can be approximated as sitting on a locally constant baseline, allowing for a rough comparison of their relative strengths. However, the true strengths cannot be accurately determined because they depend on the absolute continuum level, which is difficult to establish. In the following sections, we examine subsets of our identified features that are also detected in a separate dataset or reported in independent studies. These shared detections allow us to compare CH$_4$ absorption strengths across different sources.

The wavelength of each identified feature is assigned to the peak position in the Titan intrinsic spectrum, with an uncertainty spanning three wavelength pixels centered on that peak. In UHR mode, ESPRESSO samples one resolution element with approximately 2.5 pixels. Accordingly, we adopt a conservative three-pixel span as the uncertainty in peak position, corresponding to approximately 0.027~\AA\ at the bluest detected features and up to 0.039~\AA\ at the red end of the spectrum. A more precise estimate is not pursued, as the wings of many absorption features appear to be distorted, presumably due to blending or noise, particularly in the weaker features. Clean absorption profiles, similar to those observed in solar lines within the same Titan spectra, are rare and generally appear only in strong, isolated features. 

\section{Results} 
\label{sec:results}

We identified 6,195 absorption features within the 5400–7870~\AA\ range, potentially associated with CH$_4$ absorption. Among these, 1,207 features lie near strong solar or telluric lines, making them less reliable as genuine detections. A comprehensive list of these features, including their wavelength positions and relative strengths, is provided in Table~\ref{tab:ch4_candidates}.

\begin{table*}
\centering
\caption{Identified CH$_4$ absorption features from the ESPRESSO data, with wavelengths provided in vacuum. Relative strengths, ranging from 0 to 1, represent the absorption fraction relative to the best-fit background. The SNR is calculated as the ratio of feature strength to the local noise, measured within a 2 \AA\ window centered on the feature. The wavelength and relative strength uncertainty columns denote the $\pm$ uncertainties associated with each corresponding measurement. UVES columns with available data indicate features that have corresponding UVES detections. Columns marked with ‘Yes’ or ‘No’ indicate specific feature characteristics, including matches with RS24, SO95, CP23's Kitt Peak (KP) and CRDS data, as well as proximity to strong stellar or telluric lines (ST Mask), as detailed in the text. The features presented here serve as a representative sample covering all identified characteristics, with the complete dataset available in machine-readable format. A portion is displayed here for reference regarding its structure and content.}
\label{tab:ch4_candidates}
\begin{tabular}{c|c|c|c|c|c|c|c|c|c|c|c}
\hline
\multicolumn{5}{c|}{ESPRESSO} & \multicolumn{2}{c|}{UVES} & RS24 & SO95 & \multicolumn{2}{c|}{CP23} & ST \\
\cline{1-5} \cline{6-7} \cline{10-11}
\multicolumn{1}{c|}{Wl} & 
\multicolumn{1}{c|}{Wl Err} & 
\multicolumn{1}{c|}{Rel Str} & 
\multicolumn{1}{c|}{Rel Str Err} & 
\multicolumn{1}{c|}{SNR} & 
\multicolumn{1}{c|}{Wl} & 
\multicolumn{1}{c|}{Rel Str} & 
\multicolumn{1}{c|}{ } & 
\multicolumn{1}{c|}{ } & 
\multicolumn{1}{c|}{KP} & 
\multicolumn{1}{c|}{CRDS} & 
\multicolumn{1}{c}{Mask} \\
(\AA) & (\AA) & & & & (\AA) & & & & & & \\
\hline\hline
5422.107 & 0.027 & 0.035 & 0.006 & 6 & 5422.111 & 0.018 & N & N & N & N & Y \\
6099.144 & 0.031 & 0.023 & 0.004 & 6 & 6099.139 & 0.010 & N & N & N & N & N \\
6111.536 & 0.031 & 0.047 & 0.003 & 10 & 6111.550 & 0.034 & Y & N & N & N & N \\
6147.212 & 0.031 & 0.075 & 0.001 & 19 & 6147.202 & 0.043 & Y & N & N & N & N \\
6149.088 & 0.031 & 0.071 & 0.003 & 16 & 6149.073 & 0.046 & Y & N & N & N & N \\
6161.027 & 0.031 & 0.045 & 0.006 & 9 & - & - & N & N & N & N & N \\
6176.809 & 0.031 & 0.089 & 0.011 & 18 & 6176.794 & 0.059 & Y & N & N & N & Y \\
6180.921 & 0.031 & 0.051 & 0.003 & 11 & 6180.920 & 0.037 & Y & N & N & N & N \\
6184.736 & 0.031 & 0.125 & 0.004 & 30 & 6184.740 & 0.102 & Y & N & N & N & N \\
6191.547 & 0.031 & 0.087 & 0.005 & 19 & 6191.551 & 0.052 & N & N & N & N & N \\
6194.625 & 0.031 & 0.038 & 0.008 & 7 & - & - & N & N & N & N & N \\
6198.253 & 0.031 & 0.062 & 0.013 & 9 & 6198.255 & 0.033 & N & N & N & N & N \\
6207.139 & 0.031 & 0.094 & 0.006 & 22 & 6207.137 & 0.055 & N & N & N & N & N \\
6510.870 & 0.033 & 0.029 & 0.003 & 7 & - & - & N & N & N & N & N \\
6661.578 & 0.033 & 0.041 & 0.005 & 8 & - & - & N & N & N & N & N \\
7004.161 & 0.035 & 0.073 & 0.003 & 15 & - & - & N & N & N & N & N \\
7168.176 & 0.036 & 0.079 & 0.011 & 12 & - & - & N & N & N & N & Y \\
7218.189 & 0.036 & 0.171 & 0.004 & 26 & - & - & N & N & N & N & Y \\
7231.142 & 0.036 & 0.038 & 0.008 & 6 & - & - & N & N & Y & N & N \\
7253.947 & 0.036 & 0.254 & 0.004 & 31 & - & - & N & Y & N & N & Y \\
7256.198 & 0.036 & 0.129 & 0.008 & 14 & - & - & N & N & N & N & Y \\
7257.299 & 0.036 & 0.294 & 0.002 & 38 & - & - & N & Y & Y & N & N \\
7260.156 & 0.036 & 0.325 & 0.002 & 46 & - & - & N & Y & Y & N & Y \\
7260.725 & 0.036 & 0.323 & 0.003 & 46 & - & - & N & Y & Y & N & Y \\
7281.850 & 0.036 & 0.245 & 0.013 & 24 & - & - & N & Y & N & N & N \\
7295.732 & 0.037 & 0.244 & 0.007 & 44 & - & - & N & Y & N & N & N \\
7299.894 & 0.037 & 0.228 & 0.005 & 47 & - & - & N & Y & N & N & N \\
7386.810 & 0.037 & 0.062 & 0.012 & 6 & - & - & N & N & N & N & N \\
7526.047 & 0.038 & 0.056 & 0.005 & 10 & - & - & N & N & N & Y & N \\
7533.268 & 0.038 & 0.024 & 0.007 & 5 & - & - & N & N & N & Y & Y \\
7537.390 & 0.038 & 0.030 & 0.004 & 7 & - & - & N & N & N & Y & N \\
7549.681 & 0.038 & 0.035 & 0.001 & 8 & - & - & N & N & N & Y & N \\
7579.176 & 0.038 & 0.057 & 0.008 & 12 & - & - & N & N & N & Y & N \\
7799.412 & 0.039 & 0.137 & 0.003 & 18 & - & -  & N & N & N & N & Y \\
7821.348 & 0.039 & 0.187 & 0.008 & 31 & - & - & N & N & N & N & N \\
7855.100 & 0.039 & 0.225 & 0.013 & 28 & - & - & N & N & Y & N & N \\
7869.773 & 0.039 & 0.110 & 0.020 & 6 & - & - & N & N & N & N & N \\
\hline
\end{tabular}
\end{table*}

\begin{figure*}[ht]
    \centering
    \includegraphics[width=\textwidth]{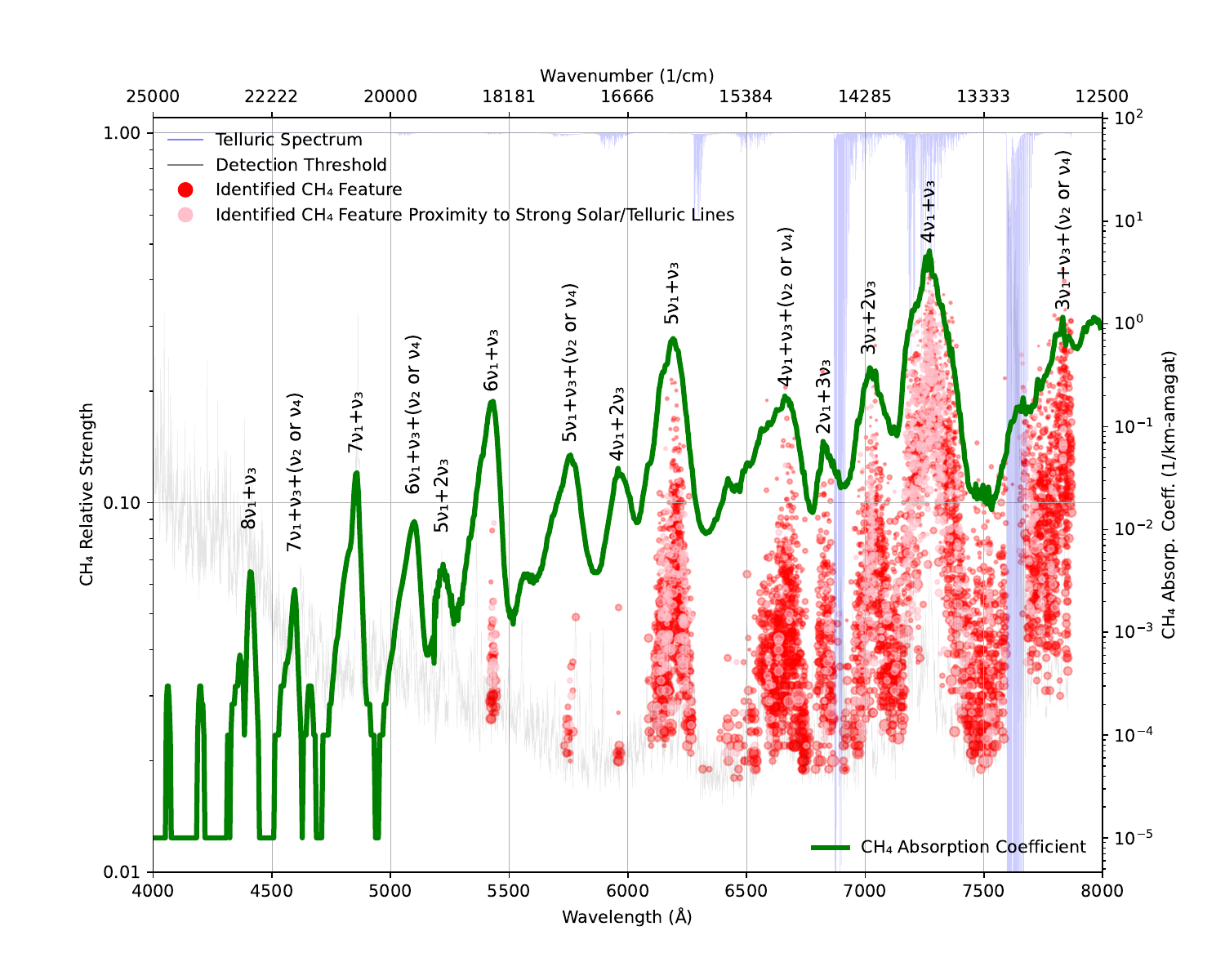} \caption{Identified CH$_4$ features, with their relative strengths plotted against vacuum wavelengths and wavenumbers. The green line represents CH$_4$ absorption coefficients at 100 K from \cite{KARKOSCHKA2010674}, serving as a reference for the locations and strengths of absorption bands. The left vertical axis corresponds to the relative strengths of the absorption features identified in this work, while the right vertical axis represents the absorption coefficients from \cite{KARKOSCHKA2010674}. The gray line indicates the detection threshold applied during the line identification process (5 times the local noise level). The blue line represents telluric absorption across different wavelengths. The size of the circles is proportional to the ratio of strength uncertainty to the measured strength. Vibrational band assignments, based on \cite{GIVER1978311}, are annotated in black text.}
    \label{fig:ch4_candidates}
\end{figure*}

Figure~\ref{fig:ch4_candidates} shows the relative strengths of the identified features plotted against vacuum wavelength and wavenumber. For reference, CH$_4$ absorption coefficients at 100 K from \citet{KARKOSCHKA2010674} and vibrational band assignments from \citet{GIVER1978311} are also included. The left vertical axis indicates our relative strength measurements, while the right vertical axis corresponds to the CH$_4$ absorption coefficients. Although the two axes are not directly proportional, the plot reveals a clear trend: the identified features tend to cluster around the major CH$_4$ absorption bands reported in the literature.

In this study, we analyzed Titan's spectra over the 4000–7870~\AA\ wavelength range. However, the bluest CH$_4$ feature identified was located near 5400~\AA, with several factors limiting detectability at shorter wavelengths. One possible contributing factor is Titan’s thick atmospheric haze, which becomes increasingly opaque below $\sim$6000~\AA\ and significantly alters the observed spectrum by modifying the continuum slope through wavelength-dependent scattering and absorption \citep{MCKAY2001}. In this region, radiative transfer is dominated by fractal haze particles whose strong scattering properties reduce the contrast of molecular absorption lines (e.g., \citealt{Batalha_2019}; \citealt{2024Rukdee}). As a result, CH$_4$ features appear weaker than they would in a clear atmosphere.

Beyond the influence of haze, CH$_4$ feature detection at shorter wavelengths is further constrained by a combination of spectral and observational challenges. Below 4500~\AA, strong and closely spaced solar lines make it difficult to distinguish them from intrinsic Titan features. Moreover, the SNR in this region is relatively low, averaging below 100, and CH$_4$ absorption is intrinsically weak, with absorption coefficients $\lesssim$ 0.001~km-amagat$^{-1}$. Around 4850~\AA\ (7$\nu_1$+$\nu_3$), the density of strong solar lines decreases, the SNR improves by a factor of two, and CH$_4$ absorption coefficients increase by an order of magnitude. Despite these improvements, CH$_4$ features remain undetectable, most likely due to limited precision in the continuum placement introduced by the broad and strong H-$\beta$ absorption line. To explore the potential for detection in this region, we tested a relaxed detection criterion by lowering the required absorption strength from five to three times the local noise level. This adjustment led to the identification of a few candidate features, though large uncertainties in absorption strength make it unclear whether these represent genuine detections or noise. In contrast, between 5700 and 6000~\AA\ (5$\nu_1$+$\nu_3$+($\nu_2$ or $\nu_4$) and 4$\nu_1$+2$\nu_3$), where CH$_4$ absorption coefficients are comparable to those around 4850~\AA, CH$_4$ features were successfully detected. This success is attributed to the significantly higher SNR in this part of the spectrum, approximately 350, along with reduced contamination from strong and closely spaced solar lines, resulting in a more stable continuum.

We estimate the detection limit for CH\textsubscript{4} features to correspond to an absorption coefficient of around 0.02~km-amagat$^{-1}$. This value is determined from the lowest CH\textsubscript{4} absorption coefficients reported in \cite{KARKOSCHKA2010674} at which corresponding features in our data could still be identified. These features appear around 7530 \AA\ under the most favorable conditions, where the local SNR of the Titan spectrum exceeds 300, the variation among the three Titan spectra is approximately 0.5\%, solar and telluric contamination is minimal, and the continuum remains stable and free of anomalies. This detection limit is a conservative estimate. As shown in the following sections, comparisons with linelists from other studies reveal several features common to both our dataset and theirs that correspond to absorption coefficients as low as 0.001 km-amagat$^{-1}$, such as those illustrated in the lower right panel of Figure \ref{fig:6examples}. However, these features are detected with lower SNRs and appear in highly congested spectral regions, which reduce confidence in their individual identification.

\section{Impact of Spectral Resolution on Methane Features} 
\label{sec:resolutions}

While retrieving ESPRESSO Titan data from the archive, we came across Titan spectra obtained with the UVES (R $\approx$ 110,000; $\Delta\lambda$ = 4143–6277~\AA; \citealt{2000SPIE.4008..534D}). These spectra provide an opportunity to examine how CH$_4$ features vary with spectral resolution. Therefore, we analyze the UVES Titan spectra as a complementary dataset to the ESPRESSO data.

UVES is a high-resolution echelle spectrograph mounted on the VLT. The UVES Titan spectra used in this study were obtained under program ID 267.C-5701(A) (PI: Courtin, R.) between 2002 and 2003. A key advantage of these data is their multi-epoch nature, spanning several months. This extended timeframe introduces significant Doppler shifts among solar, telluric, and Titan’s intrinsic features, aiding in their differentiation. We excluded spectra with significantly low SNR, resulting in a final selection of 15 UVES spectra with SNRs greater than 138. Details of the UVES spectra are provided in Table~\ref{tab:observations}. Since the Doppler shifts of spectral features observed within a single night are minimal, we combined spectra acquired on the same night to enhance the SNR, following the same approach used for the ESPRESSO data, except for a single spectrum obtained on September 9. This process resulted in five combined UVES spectral sets and one additional spectrum, covering six separate nights over four months.   
We identify potential CH$_4$ features in the UVES spectra using methods slightly modified from those applied to the ESPRESSO data. This is because our experiments indicate that, given the same detection threshold, UVES does not reveal any features absent in ESPRESSO, primarily due to UVES’s lower resolution and comparable SNR. Additionally, the three types of spectral features, including solar, telluric, and Titan’s intrinsic features, experience significant Doppler shifts, facilitating their distinction. For each night of Titan observations, we generate a best-fit background spectrum corresponding to that night's Titan spectrum and divide the Titan spectrum by this background to obtain a residual spectrum. This process yields one residual spectrum per night, resulting in a total of six residual spectra. We then combine the six residual spectra into a single residual spectrum, referred to as the UVES Titan intrinsic spectrum, which is included as supplementary data in Appendix \ref{sec:appendixa}. In this combined spectrum, Titan's intrinsic features remain aligned, while solar and telluric features do not align due to their varying Doppler shifts. We identify potential CH$_4$ features in the UVES residual spectrum when their strength exceeds three times the standard deviation of the six residual spectra. Notably, we lower the detection threshold from five in ESPRESSO to three in UVES, as we expect the strength of UVES features to be lower due to its lower spectral resolution, while both UVES and ESPRESSO spectra have similar noise levels.

\begin{figure*}[htbp]
    \centering
    \includegraphics[width=\textwidth]{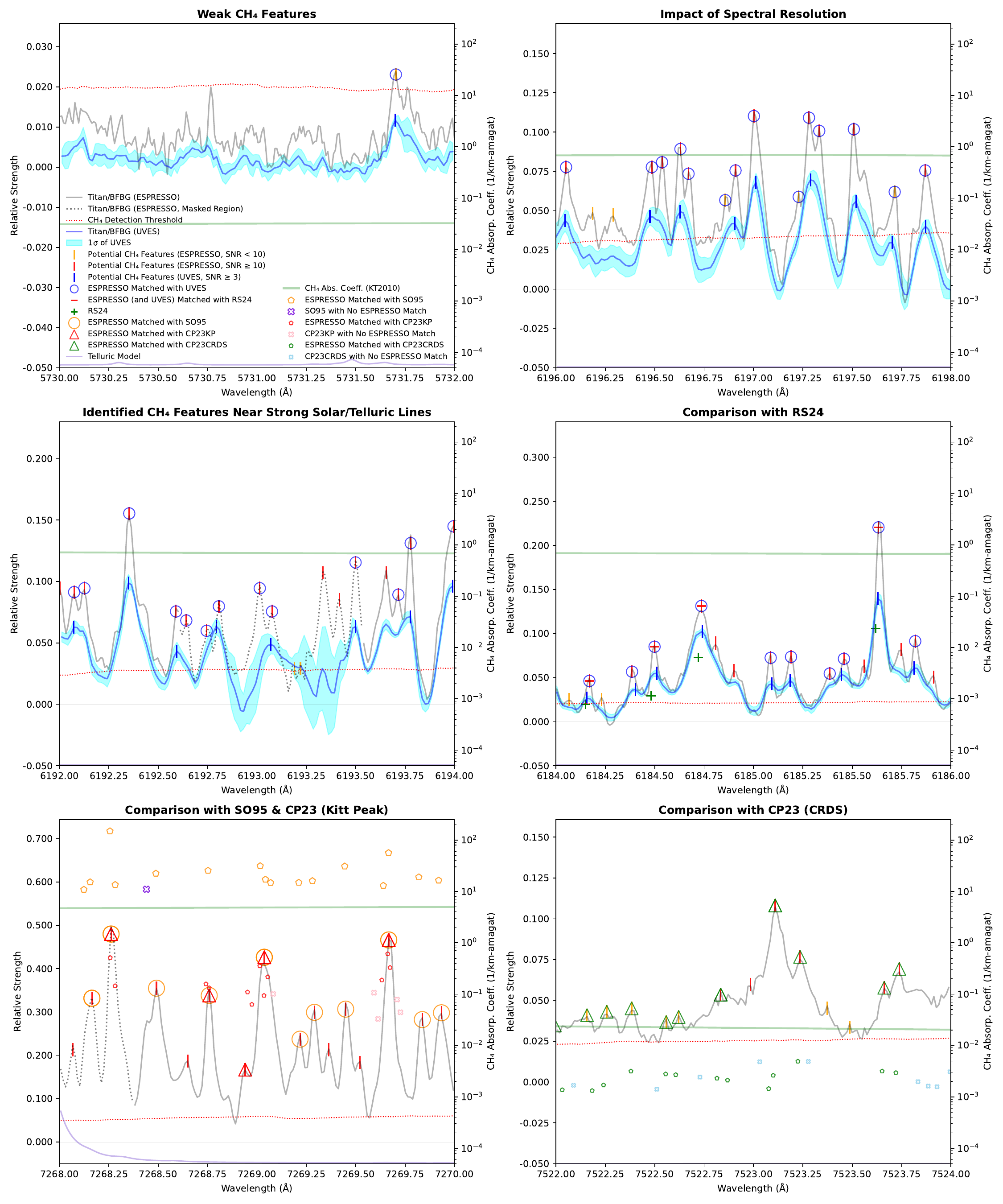} \caption{Six selected regions illustrate key aspects of the CH$_4$ features identified in this study, as described in the main text. Line plots of the normalizd ESPRESSO and UVES Titan spectra, divided by their respective best-fit backgrounds (BFBG), are shown in an inverted format to emphasize absorption strength. The telluric model is also plotted in an inverted format, with its strength scaled to fit the vertical axis. In this scale, the lowest value represents no absorption, and the highest corresponds to 100\% absorption. The legend on the left side refers to the left vertical axis, while the legend on the right refers to the right axis. A key for all symbols used in the plots is provided in the top-left panel.}
    \label{fig:6examples}
\end{figure*}

The top and middle panels of Figure~\ref{fig:6examples} show the Titan intrinsic spectra from ESPRESSO (gray) and UVES (blue), plotted in inverted format to highlight absorption strength. The gray dotted regions in the ESPRESSO spectrum indicate areas where strong solar or telluric components are present. The light blue shading represents the 1$\sigma$ region of the UVES residual spectra, where areas with noticeably broader shading (e.g., $\sim$6193.25~\AA) indicate misalignment among individual UVES residual spectra due to shifts in solar or telluric lines. In the plots, these areas are consistent with regions of strong solar or telluric components in the ESPRESSO spectrum. The red dotted line and the green line are provided as references, indicating the detection threshold for CH$_4$ feature identification in the ESPRESSO spectrum and the CH$_4$ absorption coefficient from \cite{KARKOSCHKA2010674}, respectively.

The six wavelength ranges shown in Figure~\ref{fig:6examples} were selected to illustrate key aspects of the identified CH$_4$ features and are not arranged in ascending wavelength order. The top-left panel highlights a region with relatively weak CH$_4$ features, with one CH$_4$ feature identified in the ESPRESSO spectrum and confirmed using UVES data, marked with a blue circle. The spectral shapes from both instruments show reasonable agreement. It is unclear whether visible peaks in the ESPRESSO spectrum result from genuine CH$_4$ features due to its higher resolving power or are simply noise. Several features that appear to be genuine CH$_4$ absorptions are visible, but their confirmation is challenging. The absence of detection in most peaks within this region is attributed to our conservative detection threshold, which prioritizes minimizing false positives over completeness. Detecting such weak features may require spectra with a significantly higher SNR and resolution.

The top-right panel of Figure~\ref{fig:6examples} illustrates a region with significantly stronger CH$_4$ features compared to the previously discussed region. In this region, all strong ESPRESSO CH$_4$ features (SNR $\geq$ 10) align well with their corresponding features in the UVES spectrum, while weaker features (SNR $<$ 10) are occasionally absent in the UVES data. Some features that appear blended in the UVES spectrum are partially resolved in the ESPRESSO spectrum. The plot also suggests that at the resolving power of ESPRESSO, there are CH$_4$ lines that our technique is not able to resolve (e.g. $\sim$6197.6 \AA). The middle-left panel of Figure~\ref{fig:6examples} highlights a region, part of which is located near strong solar lines. In this region, some ESPRESSO CH$_4$ features are identified but not confirmed by the UVES spectrum (e.g. between 6193.2 to 6193.4 \AA). Typically, as shown in Figure~\ref{fig:titan_bg}, absorption artifacts near strong solar or telluric lines tend to be weak. If the identified CH$_4$ features near those lines are particularly strong, they are likely genuine. However, we apply a cautionary mask to CH$_4$ features detected in these regions to encourage careful interpretation, particularly in areas prone to strong telluric contamination.
 
In the wavelength range where the ESPRESSO and UVES spectra overlap, 695 CH$_4$ features were identified in the ESPRESSO spectrum. Among these, 440 were matched to 383 CH$_4$ features detected in the UVES spectrum. The results are presented in Table~\ref{tab:ch4_candidates}, which lists the wavelength and relative strength of each UVES feature corresponding to an ESPRESSO feature, where a single UVES feature may correspond to multiple ESPRESSO features. Note that in this study, feature matching between different datasets is performed by searching, for a given feature, within a window defined by the sum of the wavelength uncertainties from both sources. For UVES, we adopt an uncertainty of two wavelength pixels, corresponding to the number of pixels used to sample one spectral resolution element. The relative strength ratio of CH$_4$ features common to both ESPRESSO and UVES is 1.512 $\pm$ 0.332, accounting for cases where a single UVES feature matches multiple ESPRESSO features. Comparing the relative strength of CH$_4$ features in ESPRESSO, those that match UVES detections tend to be stronger than those that do not. On average, the matched features are approximately 40\% stronger than the unmatched ones. These findings provide insights into how CH$_4$ absorption features vary with the resolving power of the instrument used to record the spectra.

In addition, we use the Pearson correlation coefficient to assess the linear relationship between feature strengths across dataset pairs. This serves as an indicator of the consistency in strength measurements across different sources and may help users estimate the degree of deviation when applying our results in practical contexts that differ from the specific data or methodology used in this study. The Pearson coefficient ranges from –1 to 1, with values close to 1 indicating a strong positive linear correlation, values near 0 indicating little to no linear correlation, and values close to –1 indicating a strong negative linear correlation. Table~\ref{tab:compare_strength} summarizes the computed Pearson correlation coefficients, along with notes on key factors that may contribute to deviations from perfect linearity. The corresponding p-values are all well below 0.001 and are therefore omitted from the table. When comparing datasets with different spectral resolutions, a single feature in the lower-resolution data may correspond to multiple features in the higher-resolution dataset. In such cases, the closest match in wavelength is selected for the calculation. Including all matched features instead of selecting only the closest one tends to slightly reduce the correlation coefficient. For the correlation calculated from the relative strengths of features in the ESPRESSO and UVES spectra, the Pearson coefficient is 0.97, indicating a high degree of consistency in our measured relative strengths across the two instruments.

\begin{table*}
\centering \caption{Pearson correlation coefficients assessing the consistency of methane feature strengths across dataset pairs. Values in parentheses indicate the coefficients calculated from the subset of features common to all three datasets: ESPRESSO, SO95, and CP23KP.}
\label{tab:compare_strength}
\renewcommand{\arraystretch}{1.3} \begin{tabular}{l|c|l} 
\hline
\multicolumn{1}{c|}{\textbf{Feature Pair}} & 
\multicolumn{1}{c|}{\textbf{Pearson Coefficient}} & 
\multicolumn{1}{c}{\textbf{Remarks}} \\
\hline\hline
ESPRESSO and UVES (relative strength) & 0.97 & Different spectral resolutions \\[0.8ex]
UVES and RS24 (relative strength) & 0.93 & Different in continuum placements \\[0.8ex]
ESPRESSO (relative strength) and & 0.77 (0.78)& Different spectral resolutions \\[-0.8ex]
\hspace{1.0ex} SO95 (log line intensity) &  &  \\[0.8ex]
ESPRESSO (relative strength) and & 0.56 (0.50)& Different spectral resolutions and temperatures \\[-0.8ex]
\hspace{1.0ex} CP23KP (log line intensity) &  &  \\
SO95 and CP23KP (line intensity) & (0.53) & Different temperatures \\
ESPRESSO (relative strength) and & 0.55 & Different spectral resolutions and temperatures \\[-0.8ex]
\hspace{1.0ex} CP23CRDS (log line intensity) &  &  \\
ESPRESSO (relative strength) and & 0.06 & Observation vs. \textit{ab initio} \\[-0.8ex]
\hspace{1.0ex} TheoReTS (log line intensity) &  &  \\
ESPRESSO (relative strength) and & 0.15 & Observation vs. \textit{ab initio}, special matching method \\[-0.8ex]
\hspace{1.0ex} TheoReTS (log line intensity) &  & \\
\hline
\end{tabular}
\end{table*}

\section{Comparison with Previous Studies} \label{sec:compare}

\subsection{Rianço-Silva et al.} \label{subsec:RS24}
\cite{RIANCOSILVA2024105836}, hereafter RS24, also analyzed Titan spectra obtained with VLT-UVES. One of their primary objectives was to investigate CH$_4$ absorption features in wavelength regions where no high-resolution linelist is currently available. Their dataset consists of disk-averaged spectra collected over four nights within an eight-night period in 2005. Since their data are also disk-averaged, like ours, their results provide a relevant basis for comparison. To identify CH$_4$ absorption features, RS24 employed a Doppler-based method to distinguish Titan’s atmospheric features from backscattered solar lines by leveraging Doppler shift variations observed across multiple nights. Intrinsic Titan features were aligned using ephemeris data. For each feature, two selection criteria were applied: a maximum allowable wavelength offset of 25 m\AA\ and a minimum relative depth of 2\%. Using these thresholds, they successfully identified 97 non-solar absorption features.

We cross-match these features with our candidate CH$_4$ features from the UVES data and find that 96 of the 97 features have corresponding entries in our list. To indicate these matches, we provide a ‘Y’ or ‘N’ flag in the ‘RS24’ column of Table~\ref{tab:ch4_candidates}. The only unmatched feature is likely a genuine intrinsic Titan feature but falls just below our detection threshold, resulting in its exclusion from our list. However, many of our detected features are not present in the RS24 list. Within the wavelength range covered by the RS24, we identified 274 CH$_4$ features in the UVES spectra, all of which were confirmed by the ESPRESSO data. Of these, 178 features are absent from the RS24 list. The CH$_4$ features absent from the RS24 list are approximately 25\% weaker than those in the common set. The middle-right panel of Figure~\ref{fig:6examples} illustrates a representative region where our identified CH$_4$ features match those in RS24. Matched features are marked by a `--' symbol, which, when overlapping with the `$\mid$' symbol indicating ESPRESSO detections and confirmed by our UVES data, appears as `$\oplus$'. In this plot, four of our features match those in RS24, while the others are absent from their list. Additionally, we provide a green `+' symbol to represent RS24's measured values, which can be directly compared with our UVES measurements, marked by a blue `$\mid$'. Notably, RS24 measurements appear weaker than our corresponding feature strengths. Among the 96 features matched to RS24, the feature strength ratio between our measurements and those in RS24 is 1.314 $\pm$ 0.314. The difference in feature strengths is most likely due to variations in the continuum levels from which the feature strengths are measured. We compute the Pearson correlation coefficient between our UVES data and the RS24 data, yielding a value of 0.93. Furthermore, we find that the wavelengths of our features are systematically 0.013~\AA\ greater than those reported in RS24. This offset is comparable to the wavelength sampling of UVES.

\subsection{Singh and O'Brien} \label{subsec:SO95}
\citet{SINGH1995607,1996Ap&SS.236...97S}, hereafter SO95, employed intracavity laser spectroscopy (ILS) to measure CH\textsubscript{4} absorption in the 727 nm band at 77 K. Among their results is a list of 316 strong CH\textsubscript{4} lines with absorption coefficients exceeding 10 km-amagat\textsuperscript{$-1$}, and line position accuracies of 0.02 cm\textsuperscript{$-1$} ($\approx$0.01~\AA). Since these measurements were conducted at a temperature comparable to that of Titan’s atmosphere, we compare their linelist with the CH\textsubscript{4} features identified in our study.

We find that 298 out of the 316 SO95 lines match 264 of our ESPRESSO features. An example of these matches is shown in the lower-left panel of Figure~\ref{fig:6examples}. In this panel, CH\textsubscript{4} features from our dataset that match SO95 lines are marked with orange circles. The corresponding SO95 lines are shown as orange polygons, while unmatched SO95 lines appear as purple crosses. Since SO95 reported only strong lines, many of the features we identified do not have corresponding matches in their list. Conversely, 18 SO95 lines do not match any of our detected features. These unmatched lines can be attributed either to blending with nearby features, which makes them undetectable in our data (e.g., at $\sim$7268.40~\AA), or to their location within regions affected by strong solar or telluric contamination, where only faint signs of a match are visible upon visual inspection.

Because the SO95 data were acquired at significantly higher spectral resolution ($R \geq 500{,}000$), multiple SO95 lines can correspond to a single ESPRESSO feature. For example, near 7269.00~\AA, three SO95 lines match a single feature in our dataset. As described previously, a ``match” in this context means that, for a given line in one dataset, one or more lines in the other dataset fall within a matching window defined by the combined uncertainties of both datasets. This does not imply that the matched lines corresponds to the same quantum transition. However, the match closest in wavelength is more likely to originate from the same transition, although this is not guaranteed. In lower-resolution data, the apparent position of a feature can shift noticeably from its higher-resolution counterparts (e.g., at $\sim$6197.50~\AA). 

The wavelength difference between each feature in our dataset and the nearest matched SO95 line is approximately 0.007~\AA, which is smaller than the wavelength uncertainty reported for the SO95 line positions. This indicates good agreement between the two lists. We also compute the Pearson correlation coefficient between the relative strengths of our identified features and the logarithm of the absorption coefficients reported by SO95, yielding a value of 0.77 (Table~\ref{tab:compare_strength}). This reflects a moderately strong positive linear relationship, suggesting that stronger features in our dataset generally correspond to higher absorption coefficients, consistent with physical expectations. This level of correlation is reasonable, given that our measurements are relative rather than absolute, the SO95 values approximate absolute line strengths, and many features in our dataset are blended.

\subsection{Campargue et al.} 
\label{subsec:CP23}
\cite{Campargue_2023}, hereafter CP23, conducted a detailed study of methane absorption in the \(\text{10800--14000 cm}^{-1}\) ($\sim$7140--9260 \AA) range using archival Kitt Peak FTS and high-sensitivity Cavity Ring-Down Spectroscopy (CRDS) data, both recorded at room temperature ($\sim$298 K). Their results provide a valuable reference for assessing the temperature dependence of CH\textsubscript{4} absorption in this spectral region. 

Their linelist includes approximately 12,800 lines from the FTS, with line position uncertainties of 0.005 cm\textsuperscript{–1}, and an additional 1,792 lines from the CRDS, with uncertainties of 0.001 cm\textsuperscript{–1}, covering the narrower 13060–13300 cm\textsuperscript{–1} (
$\sim$7515–7655~\AA) range. A key finding was the detection of quasi-continuum absorption near 11200 cm\textsuperscript{$-1$} ($\sim$8920 \AA) and 13800 cm\(^{-1}\) ($\sim$7250 \AA), attributed to numerous unresolved weak lines. CP23 compared their results with the HITRAN and TheoReTS databases. They found good agreement with HITRAN line positions in the 10800–11502 cm\(^{-1}\) ($\sim$8695–9260 \AA) range, where the HITRAN resolution is approximately 0.01 cm\textsuperscript{–1}, although some discrepancies in line intensities were observed. HITRAN provides CH\textsubscript{4} line parameters at 296 K intended for Earth’s atmosphere, but these values are not based on direct measurements at that temperature. Instead, they are extrapolated from low-temperature data (99–161 K) and include only partial ro-vibrational assignments. In contrast, TheoReTS showed significant deviations in line positions, with some strong lines differing by as much as 2 cm\(^{-1}\). 

CP23 identified over 12,800 CH\textsubscript{4} lines from Kitt Peak FTS data (hereafter CP23KP), of which only 954 lie within the wavelength range covered by our data, primarily within the 4$\nu_1$+$\nu_3$ and 3$\nu_1$+$\nu_3$+($\nu_2$ or $\nu_4$) bands. Cross-matching their dataset with ours, we find that 737 CP23KP lines correspond to 352 features in the ESPRESSO data, with up to four CP23KP lines matching a single ESPRESSO feature. Examples of both matched and unmatched features are shown in the lower left panel of Figure~\ref{fig:6examples}. In this panel, CH$_4$ features from our dataset  that match CP23KP lines are marked with red triangles. The corresponding CP23KP lines are shown as red polygons, while unmatched CP23KP lines are indicated by pink crosses. Note that CP23 provides line intensities in units of cm per molecule. We convert these to km-amagat$^{-1}$ using the conversion factor 1 km-amagat$^{-1}$ = 3.25 $\times$ 10$^{-25}$ cm per molecule, as given in CP23. We observe that some features in our dataset are not present in CP23KP (e.g., $\sim$7269.20–7269.50~\AA), while some CP23KP lines are missing from our data (e.g., $\sim$7269.60~\AA). In the latter case, some lines show only faint traces and remain unidentified in the ESPRESSO spectrum, while others are entirely absent.

The Pearson correlation coefficient between the relative strengths of ESPRESSO features and the logarithm of the line intensities reported in CP23KP is 0.56 (Table \ref{tab:compare_strength}). The CP23KP intensities reflect the line component measured above the quasi-continuum absorption, which is produced by unresolved weak lines, and may serve as a rough approximation of relative line strengths. However, their continuum levels may differ substantially from ours, due in part to differences in measurement techniques and temperature conditions.

Within the 7186–7370~\AA\ range, 141 features are common to the ESPRESSO, SO95, and CP23KP datasets, allowing for a more detailed comparison of feature strengths. For this subset, we compute the Pearson correlation coefficients between each pair of the datasets, with the results reported in parentheses in Table~\ref{tab:compare_strength}. The correlation coefficients for dataset pairs that differ in temperature are significantly lower. Together with the results shown in Figure~\ref{fig:6examples}, these findings indicate that CH\textsubscript{4} absorption clearly varies with temperature, affecting both the strength and appearance of individual lines.

Of the 1,792 CH$_4$ lines from the CRDS (hereafter CP23CRDS), only 244 match our 180 features. Examples of these matches are shown in the bottom-right panel of Figure~\ref{fig:6examples}. In this panel, CH\textsubscript{4} features from our dataset that match CP23CRDS lines are marked with green triangles. The corresponding CP23CRDS lines are shown as green polygons, while unmatched CP23CRDS lines are indicated by light blue crosses. Beyond temperature dependence, two main factors contribute to the limited number of matches. First, approximately 50\% of the CP23CRDS lines lie within the 7596–7655~\AA\ range, which is heavily affected by telluric absorption. Second, CH\textsubscript{4} features in this region are highly congested, resulting in low peak prominence and, consequently, missed identifications in our dataset. However, comparing the ESPRESSO spectral shape to the locations of CP23CRDS lines suggests that many of these features are not entirely absent, but rather marginally undetected by our pipeline (e.g., $\sim$7522.75, 7523.90~\AA). As with CP23KP, we also detect features that are not present in CP23CRDS (e.g., $\sim$7523.50 \AA). The Pearson correlation coefficient between the relative strengths of our CH\textsubscript{4} features and the logarithm of CP23CRDS line intensities is 0.55 (Table \ref{tab:compare_strength}), consistent with previous results in cases where temperature differences exist between datasets. We include `Y' or `N' flags in Table~\ref{tab:ch4_candidates} under the columns `CP23KP' and `CP23CRDS' to indicate whether a match is present or absent, respectively.

\subsection{TheoReTS} 
\label{subsec:TheoReTS}
TheoReTS (Theoretical Reims-Tomsk Spectral data; \citealt{REY2016138}) is an online database that provides linelists for several molecular species, including CH\textsubscript{4}, based on \textit{ab initio} calculations supplemented with experimental refinements. For CH\textsubscript{4}, linelists are available at two temperatures: 80 K (low) and 296 K (room temperature). We compared our identified CH\textsubscript{4} features with the TheoReTS linelist at 80~K\footnote{\url{https://theorets.tsu.ru/molecules.ch4.low-room-t}}, which matches the temperature conditions of our data. 

Within the wavelength range overlapping with the ESPRESSO data ($\sim$7450-7870 \AA), TheoReTS reports over 200{,}000 lines, which are so densely populated that nearly any of our features could coincide with a line by chance. To reduce random matches, we applied a cutoff based on our estimated detection limit of 0.02~km-amagat\textsuperscript{–1} (approximately $6.5 \times 10^{–27}$~cm per molecule), retaining only the stronger lines from the TheoReTS linelist. A total of 3,339 TheoReTS lines satisfied this criterion. Cross-matching these with our dataset resulted in 1,364 TheoReTS lines matching 665 ESPRESSO features. Although the matched wavelengths show excellent agreement ($\sim$0.0002 \AA), the corresponding line intensities do not. This contrasts with previous comparisons, such as those involving CP23KP and CP23CRDS, where a degree of consistency remained visually evident even in congested regions or under differing temperature conditions. For TheoReTS, however, the Pearson correlation coefficient is only 0.06 (Table \ref{tab:compare_strength}), indicating the lack of a linear relationship. 

Even after filtering for strong lines, certain spectral regions in the TheoReTS dataset remain highly congested, particularly toward the red end of the ESPRESSO spectrum. To assess whether the matching method affects the result, we test an alternative approach in which, instead of selecting the nearest TheoReTS line in wavelength, we select the strongest line within a defined matching window. This method yields a slightly higher Pearson correlation coefficient of 0.15 (Table \ref{tab:compare_strength}), but still indicates a weak correlation. These findings support the conclusion from CP23 that the TheoReTS linelist is not yet sufficiently accurate for high-resolution spectroscopic applications.

\section{Summary and Discussion} 
\label{sec:summary}

We analyzed high-resolution spectra of Titan obtained with the ESPRESSO and UVES spectrographs to identify and characterize methane absorption features in the 5400–7870~\AA\ wavelength range. Our goals were to determine the wavelength positions and relative strengths of CH\textsubscript{4} features and to evaluate how spectral resolution affects their detectability. The ESPRESSO spectra offer unprecedented detail, revealing thousands of Titan absorption features potentially associated with CH\textsubscript{4}. We identified 6,195 candidate CH\textsubscript{4} features, including 5,436 newly reported in this study. At ESPRESSO’s resolving power (R~$\approx$~190{,}000), most CH\textsubscript{4} absorption lines remain unresolved, so the detected features correspond to blended structures rather than isolated transitions.

To achieve reliable feature identification and minimize false positives, we adopted a conservative detection strategy that intentionally overestimates the contributions from solar and telluric lines. This facilitates a more effective separation of these components from intrinsic features in Titan’s spectrum. A detection threshold of five times the local noise level was applied to identify CH\textsubscript{4} absorption features. While this approach improves the confidence in the detections, it reduces completeness by excluding weaker CH\textsubscript{4} features, particularly those overlapping with solar or telluric lines.

A comparison between ESPRESSO and UVES (R~$\approx$~110{,}000) spectra allowed us to assess the impact of spectral resolution on CH\textsubscript{4} feature detectability. Within the overlapping wavelength range, ESPRESSO revealed approximately twice as many features as UVES, with an average feature strength 1.5 times greater, despite the more stringent detection threshold adopted for ESPRESSO (5 times the noise level, compared to 3 times for UVES). These results demonstrate that lower spectral resolution leads to weaker features and increased blending. This finding emphasizes the importance of high spectral resolution for resolving individual CH\textsubscript{4} lines, which are essential for accurate atmospheric modeling and reliable methane detection in exoplanetary atmospheres.

To assess consistency and potential discrepancies in CH\textsubscript{4} feature identification, we compared our results with RS24 \citep{RIANCOSILVA2024105836}, SO95 (\citealt{SINGH1995607, 1996Ap&SS.236...97S}), CP23 \citep{Campargue_2023}, and the TheoReTS database \citep{REY2016138}. RS24 is based on Titan spectra, while SO95 derives from laboratory measurements at 77~K, closely approximating Titan’s atmospheric temperature ($\sim$100~K). TheoReTS similarly provides a linelist at 80~K based on \textit{ab initio} calculations, whereas CP23 reports line parameters measured at room temperature ($\sim$298~K). Our results show strong agreement with RS24 and SO95 but significantly weaker consistency with CP23, indicating the temperature dependence of CH\textsubscript{4} absorption, as also observed in near-infrared studies. By contrast, TheoReTS exhibits poor correspondence with our identifications, suggesting the current limitations of theoretical linelists for high-resolution spectroscopy.

We provide a comprehensive list of CH\textsubscript{4} candidate features, including their detection SNR and a set of flags that offer insights into the confidence of each feature's identification. These flags indicate proximity to strong solar or telluric lines, cross-matches with previously reported CH\textsubscript{4} features, and confirmations from multiple spectral datasets. Incorporating these confidence markers supports a more reliable interpretation of CH\textsubscript{4} absorption features. However, the absence of a cross-match with previous studies does not necessarily imply lower reliability, particularly for CH\textsubscript{4} features within $\sim$6190–7140~\AA, which are reported here for the first time. 

The relative strength values provided in our dataset offer useful information for characterizing CH\textsubscript{4} absorption features, but they should be interpreted with caution. These values are derived from normalized spectra and therefore do not represent absolute absorption strengths, but rather relative comparisons within our dataset. Several factors can influence these measurements, including uncertainties in continuum placement, spectral resolution, and the presence of quasi-continuum absorption, as described in CP23. While the relative strengths are valuable for comparative analyses, they should not be interpreted as intrinsic line strengths. To provide a quantitative perspective on their reliability, we report the Pearson correlation coefficient as an indicator of how our relative strengths deviate from a linear relationship with line strengths reported in various datasets.

Despite these limitations, the relative strengths remain valuable, particularly in high-resolution spectra, where recovering the true flux is inherently challenging due to atmospheric and instrumental effects. They are especially useful in differential analyses, where relative line strengths can be compared across different datasets or models. Additionally, our dataset can support the development of a binary mask for cross-correlation methods, which are essential for identifying CH\textsubscript{4} signatures in high-resolution spectra. Such a mask, based on our identified features, can help isolate spectral regions where methane absorption is expected, thereby improving retrieval techniques under challenging observational conditions.

This work contributes to the development of high-resolution CH\textsubscript{4} absorption data, which remains limited in the visible wavelength range. This limitation poses significant challenges for the detection and characterization of CH\textsubscript{4} in planetary and exoplanetary atmospheres when using current and next-generation high-resolution spectrographs. The dataset presented here helps address this constraint and also supports the derivation of a more complete and accurate CH\textsubscript{4} linelist based on \textit{ab initio} methods, ultimately improving the accuracy of theoretical models. These advancements enhance our ability to investigate methane as a key tracer of atmospheric processes and a potential biosignature.

\begin{acknowledgments}
We thank the anonymous referees for the constructive comments, which helped improve the clarity and quality of this work. We thank Sergey Yurchenko for helpful discussions that provided motivation and valuable references for this work. This work is based on observations collected at the European Organisation for Astronomical Research in the Southern Hemisphere under ESO programmes 106.218L.001 and 267.C-5701(A). All data are publicly available through the ESO Science Archive. The NSO/Kitt Peak FTS data used in this study were produced by NSF/NOAO. This research has made use of the  TAPAS service through AERIS center. This work is supported by the Fundamental Fund of Thailand Science Research and Innovation (TSRI) through the National Astronomical Research Institute of Thailand (Public Organization) (FFB680072/0269).
\end{acknowledgments}

\appendix
\section{Intrinsic Titan Spectra}
\label{sec:appendixa}
The intrinsic Titan spectral data referenced in Section~\ref{subsec:featuredepth} and Section~\ref{sec:resolutions} are described below. The data are provided separately for the ESPRESSO and UVES observations, and include only the spectral regions with wavelengths greater than 5400~\AA (see Tables~\ref{tab:A1} and \ref{tab:A2}, respectively).

For the ESPRESSO spectrum, the file contains four columns:
\begin{itemize}
\item Wavelength in vacuum (\AA),
\item Relative flux (unitless; normalized Titan spectrum with the best-fit background removed),
\item Relative flux uncertainty (variation in the relative flux across the three individual spectra),
\item A binary mask indicating whether the corresponding wavelength is in the vicinity of strong solar or telluric lines (0 = yes, 1 = no).
\end{itemize}

For the UVES spectra, the file contains three columns:
\begin{itemize}
\item Wavelength in vacuum (\AA),
\item Relative flux (unitless; normalized Titan spectrum with the best-fit background removed),
\item Relative flux uncertainty (variation in the relative flux across the 15 individual spectra).
\end{itemize}

The UVES data were collected over a relatively broad time window, making the flux variations sufficient to indicate wavelengths potentially affected by strong solar or telluric lines. As a result, a binary mask is not provided for the UVES data, unlike for the ESPRESSO spectra. We caution that spectral regions outside the identified CH$_4$ features may be more affected by artifacts and residual solar or telluric absorption. In contrast, the regions containing identified CH$_4$ features have undergone more careful evaluation, as described in the main text.

\begin{deluxetable}{cccc}
\tablenum{A1}
\tablecaption{Intrinsic spectral data from ESPRESSO observations\label{tab:A1}}
\tablehead{\colhead{Wavelength} & \colhead{Relative Flux} & \colhead{Relative Flux Uncertainty} & \colhead{Mask} \\ \colhead{\AA} & \nocolhead{} & \nocolhead{} & \nocolhead{}}
\startdata
5400.006&0.998&0.005&1\\
5400.015&0.999&0.006&1\\
5400.024&0.996&0.005&1\\
5400.033&0.990&0.008&1\\
5400.042&0.988&0.008&1\\
5400.052&0.996&0.009&1\\
5400.061&0.993&0.001&1\\
5400.070&0.984&0.002&1\\
5400.079&0.983&0.002&1\\
5400.088&0.986&0.003&1\\
...&...&...&...
\enddata
\tablecomments{The full version of this table is available in machine-readable format in the online journal. A portion is shown here for guidance regarding its form and function.}
\end{deluxetable}

\begin{deluxetable}{ccc}
\tablenum{A2}
\tablecaption{Intrinsic spectral data from UVES observations\label{tab:A2}}
\tablehead{\colhead{Wavelength} & \colhead{Relative Flux} & \colhead{Relative Flux Uncertainty} \\ \colhead{\AA} & \nocolhead{} & \nocolhead{}}
\startdata
5400.005&0.995&0.002\\
5400.021&0.994&0.002\\
5400.036&0.993&0.004\\
5400.051&0.993&0.005\\
5400.067&0.993&0.005\\
5400.082&0.993&0.004\\
5400.097&0.993&0.003\\
5400.113&0.994&0.004\\
5400.128&0.994&0.004\\
5400.143&0.995&0.005\\
...&...&...
\enddata
\tablecomments{The full version of this table is available in machine-readable format in the online journal. A portion is shown here for guidance regarding its form and function.}
\end{deluxetable}

\clearpage

\bibliography{opticalch4}{}
\bibliographystyle{aasjournal}

\end{document}